\documentclass[11pt]{article}    

\usepackage[margin=1in]{geometry}
\usepackage{comment}                
\usepackage[tbtags]{amsmath}        
\usepackage{xcolor}                 
\usepackage{enumitem}               
\usepackage{amssymb}                
\usepackage{amsthm}                 
\usepackage{hyperref}               
\usepackage{graphicx}               
\usepackage{tikz, pgfplots}         
\usepackage{float}                  
\usepackage{natbib}                 
\usepackage{rotating}
\usepackage{setspace}
\usepackage{booktabs}
\usepackage{cleveref}
\usepackage{tcolorbox}
\usepackage{pdfpages}
\usepackage{tikz-cd}
\tikzcdset{row sep/normal=10pt, column sep/normal=20pt}

\theoremstyle{theorem}

\newtheorem{proposition}{Proposition}
\newtheorem{lemma}{Lemma}
\newtheorem{corollary}{Corollary}
\DeclareMathOperator{\Tr}{Tr}
\numberwithin{equation}{section}

\title{%
	Forecast Relative Error Decomposition\thanks{We thank Chausse, P., and Daouia, A., for their helpful comments.}}
\author{Gouri\'eroux, C.,\thanks{University of Toronto, Toulouse School of Economics and CREST, email: Christian.Gourieroux@ensae.fr.} and Q., Lee\thanks{University of Toronto, Email: qt.lee@mail.utoronto.ca.}}
\date{\today, All comments welcome.}

\begin{document}
	\maketitle
	\begin{abstract}
		\noindent We introduce a class of relative error decomposition measures that are well-suited for the analysis of shocks in nonlinear dynamic models. They include the Forecast Relative Error Decomposition (FRED), Forecast Error Kullback Decomposition (FEKD) and Forecast Error Laplace Decomposition (FELD). These measures are favourable over the traditional Forecast Error Variance Decomposition (FEVD) because they account for nonlinear dependence in both a serial and cross-sectional sense. This is illustrated by applications to dynamic models for qualitative data, count data, stochastic volatility and cyberrisk. 	 \\

		\vspace{0in} 
	\end{abstract}
	
	\setstretch{1}

\section{Introduction}

Since its introduction in the time series literature [see e.g. Doob (1953), Whittle (1963)], the Forecast Error Variance Decomposition (FEVD) has been largely used in macroeconomics to analyse the effects of shocks and their propogation at different horizons in time [see e.g. Lanne, Nyberg (2016), Isakin, Ngo (2020)]. This approach is easy to implement at the cost of serial and cross-sectional linearity assumptions. It is typically applied on linear dynamic models such as Structural Vector Autoregressions (SVAR) with independent and identically distributed innovations. Moreover, the variance is just a local measure of uncertainty when this uncertainty is small [see Pratt (1963), Arrow (1965) for the definition of risk aversion], and their local interpretation is the basis of the short term mean-variance portfolio management in Markowitz (1953), (2000)]. The goal of our paper is to extend the standard FEVD in the two following directions above: [1] First to account for nonlinear dynamic models (nonlinear serial dependence), that involves asymmetric effects, conditional heteroscedasticity, switching regimes, or extreme risks. [2] Second, to allow not only for the prediction of the future value of the process, but also for the prediction of its nonlinear transforms (nonlinear cross-sectional dependence), in the short, medium and long run. Section 2 provides the background on the variance decomposition formula, its role in defining the standard FEVD, and the local interpretation of this decomposition. In Section 3, we extend the FEVD measure in a nonlinear dynamic framework. We first introduce a general decomposition for the relative forecast errors on a positive transformation of the process of interest, which leads to the so-called Forecast Relative Error Decomposition (FRED).  We then apply this decomposition to the updating (i.e. learning) of the forecasts and propose two decompositions based on selected characterizations of the predictive distribution: (i) The Forecast Error Kullback Decomposition (FEKD) is a measure based on the transition density. (ii) The Forecast Error Laplace Decomposition (FELD) is a measure based on the conditional Laplace transform. We demonstrate the suitability of these new decompositions for nonlinear dynamic models in Section 4. We begin with the standard linear (Gaussian) Vector Autoregression (VAR) and compare our measures with the traditional FEVD. Then, we consider different examples in which the nonlinear features are due to the structural interpretation of the process of interest, which can be qualitative, a sequence of count variables, be positive valued, or even in the case of observed volatility-covolatility features be valued as a symmetric semi-positive definite matrix. Numerical illustrations are provided in Section 5, and statistical inference is discussed in Section 6. An application to count observations on cyberattacks is provided in Section 7 and we conclude in Section 8.  Technical details are provided in the appendices and online appendices.

\section{Variance Analysis as a Local Approximation}

This section introduces two preliminary lemmas that are useful in understanding the Forecast Error Variance Decomposition and its interpretation. 

\subsection{Decomposition of Variance}

Let us consider a random vector $Y$ of dimension $n$ and an information set ($\sigma$-algebra) $I$ such that $Z_t$ is measurable with respect to $I_t$. 
\begin{lemma}[Decomposition of Variance]
If $Y$ is square integrable\footnote{that is, $\mathbb{E}\vert\vert Y \vert\vert^2<\infty$, where $\vert\vert \cdot \vert\vert$ denotes the Euclidean norm.}:
\begin{equation}\label{V1}
	\mathbb{V}[Y] = \mathbb{V}[\mathbb{E}(Y|I)] + \mathbb{E}[\mathbb{V}(Y|I)],
\end{equation}
where $\mathbb{E}(Y|I)$ (resp. $\mathbb{V}(Y|I)$) denotes the expectation (resp. variance-covariance matrix) of $Y$ conditional on the information set $I$.
\end{lemma}
This decomposition can be written as: 
\begin{equation}\label{V2}
	\mathbb{E}[(\mathbb{E}(Y|I)-\mathbb{E}(Y))(\mathbb{E}(Y|I)-\mathbb{E}(Y))']=\mathbb{V}[Y]-\mathbb{V}[\mathbb{E}(Y|I)],
\end{equation}
where $'$ denotes a transpose. It relates the multivariate risk of prediction updating (left hand side) to the updating of the variances of the error forecasts (right hand side). The matrix equations \eqref{V1}-\eqref{V2} can be decomposed by components. They lead to the one dimensional variance decompositions as:
\begin{equation*}
	\mathbb{V}(Y_i)=\mathbb{V}\left[\mathbb{E}(Y_i|I)\right] + \mathbb{E}\left[\mathbb{V}(Y_i|I)\right], \ \ i=1,...,n.
\end{equation*}
as well as one dimensional covariance decompositions:
\begin{equation*}
	Cov(Y_i,Y_j) = Cov\left[\mathbb{E}(Y_i|I),\mathbb{E}(Y_j|I)\right] + \mathbb{E}\left[Cov(Y_i,Y_j|I)\right], \ \ i,j=1,...,n, i\neq j,
\end{equation*}
in which the information set is not specific to a given component. It is easily checked that these additive decompositions of the variances and covariances do not imply a simple decomposition of the associated correlations $\rho(Y_i,Y_j)=\frac{Cov(Y_i,Y_j)}{\sqrt{\mathbb{V}(Y_i)\mathbb{V}(Y_j)}}$ in terms of the conditional correlations.

\subsection{Forecast Error Variance Decomposition}

We can extend the decomposition of variance to any multivariate stochastic process $(Y_t)$ and its increasing sequence of information sets (or filtration) $(I_t)$, where $I_t=(\underline{Y}_t)$ is the $\sigma$-algebra generated by the present and lagged values of the process. Indeed, the forecast errors at horizon $h$ can be written as a sum of multivariate forecast updates:
\begin{equation}
	\begin{split}
		 & Y_{t+h}-\mathbb{E}(Y_{t+h}|I_t) \\ 	
	= &  [Y_{t+h}-\mathbb{E}(Y_{t+h}|I_{t+h-1})]+[\mathbb{E}(Y_{t+h}|I_{t+h-1})-\mathbb{E}(Y_{t+h}|I_{t+h-2})] \\
	& + ... + [\mathbb{E}(Y_{t+h}|I_{t+1})-\mathbb{E}(Y_{t+h}|I_t)] \\
	= & \sum_{k=0}^{h-1} [\mathbb{E}(Y_{t+h}|I_{t+k+1})-\mathbb{E}(Y_{t+h}|I_{t+k})].
	\end{split}
\end{equation}
By the optimality of the conditional expectations, the forecast updatings are uncorrelated conditional on $I_t$. Then, we get the FEVD. 
\begin{lemma}[Forecast Error Variance Decomposition (FEVD)]
	\begin{equation}
		\begin{split}\label{FEVD}
			\mathbb{V}[Y_{t+h}|I_t] = & \sum_{k=0}^{h-1} \mathbb{V}\{[\mathbb{E}(Y_{t+h}|I_{t+k+1})-\mathbb{E}(Y_{t+h}|I_{t+k})]|I_t\}  \\
			= & \sum_{k=0}^{h-1} \mathbb{E}\left\{\mathbb{V}\left[\mathbb{E}\left(Y_{t+h}\vert I_{t+k+1}\right)\vert I_{t+k}\right]\vert I_t\right\}.
		\end{split}		
	\end{equation}
\end{lemma}

 This decomposition is often written for a strictly stationary process $(Y_t)$, which admits an infinite strong moving average\footnote{Since our objective is an extension of the FEVD to a nonlinear dynamic framework, it is important to consider strong moving average models, that are defined from strong white noise instead of weak moving average models where the noise is just assumed to be zero mean, with fixed variance and no serial correlation.} representation of the form:
\begin{equation}
	Y_t=\sum_{j=0}^{\infty} A_j \varepsilon_{t-j}, A_0 =  Id,
\end{equation}
where $\varepsilon_t$ is a strong white noise, that is a sequence of i.i.d random vectors, with zero mean and variance-covariance $\Sigma$, and the moving average coefficients satisfy the square integrability condition $\sum_{j=0}^{\infty} \vert\vert A_j \vert \vert^2 < \infty$. When this moving average representation is invertible, the process also admits an (infinite) autoregressive representation:
\begin{equation}
\sum_{j=0}^{\infty}\Phi_j Y_{t-j} = \varepsilon_t, \Phi_0 = Id,
\end{equation}
and the information generated by the current and lagged values of the process $(Y_t)$ is equal to the information generated by the current and lagged values of the noise $(\varepsilon_t)$, that is: 
\begin{equation}
	I_t = (\underline{Y}_t)=(\underline{\varepsilon}_t).
\end{equation}
Then, the FEVD becomes:
\begin{equation}
	\mathbb{V}\left[Y_{t+h}|\underline{Y}_t\right]=\mathbb{V}[Y_{t+h}|\underline{\varepsilon}_t]=\sum_{k=0}^{h-1} A_k\Sigma A_k',
\end{equation}
since:
\begin{equation}
	\begin{split}
	& \mathbb{E}[\mathbb{V}(Y_{t+h}|I_{t+k})|I_t] \\
	= & \mathbb{V}[\mathbb{E}(Y_{t+h}|I_{t+k+1})-\mathbb{E}(Y_{t+h}|I_{t+k})|I_t]\\
		= & \mathbb{V}[A_{h-k-1}\varepsilon_{t+k+1}|I_t]=A_{h-k-1}\Sigma A'_{h-k-1}.\\
	\end{split}
\end{equation}
In this special moving average case with iid noise, the decomposition is path independent, that is independent on the values of process $(Y_t)$ before time $t$. These decompositions in the strong moving average case, that is in the strong linear dynamic case, have been initially analyzed in a systematic way by Whittle (1963) [see also Doob (1953)].

\subsection{Local Analysis of Risk}

The comparison of multivariate quantitative risks (i.e. the losses) $X$ and $Y$ is based on the notion of stochastic dominance at order 2. $Y$ is riskier than $X$ if and only if $\mathbb{E}[\upsilon(Y)] \geq \mathbb{E}[\upsilon(X)]$, for any increasing convex function\footnote{This stochastic dominance can be equivalently defined in terms of preference, with now utility functions $u$ that are increasing concave.} $\upsilon$ (such that the expectations exist) [see Rothschild, Stiglitz (1970), Vickson (1975), Fishburn, Vickson (1978).]. In the univariate case, it is well known that the quadratic function $y \rightarrow y^2$ that underlies the definition of variance is convex, but not increasing (except if $X$ and $Y$ are nonnegative). Then the variance is not directly appropriate for measuring risk. However, we have the following local expansion. 
\begin{lemma}
	Let us assume that $\upsilon$ is increasing and convex such that $\upsilon(0)=0$, and the variable $Y$ is close to 0, with 0 mean. We have:
\begin{enumerate}[label=(\roman*)]
	\item $\mathbb{E}[\upsilon(Y)] \geq 0$.
	\item $\mathbb{E}[\upsilon(y)] \approx \frac{1}{2} \Tr\left[\frac{\partial^2 v(0)}{\partial y \partial y'}\mathbb{V}(Y)\right]$, where $\Tr$ denotes the trace operator.
\end{enumerate}
\end{lemma}
\textbf{Proof:} (i) This is a consequence of Jensen's inequality. \\

(ii) Indeed we have:
\begin{equation*}
	v(Y) \approx \frac{dv(0)}{dy'} Y + \frac{1}{2} Y'\frac{\partial^2 v(0)}{\partial y \partial y'} Y, 
\end{equation*}
and by taking the expectation of both sides:
\begin{equation*}
	\begin{split}
		\mathbb{E}\left[v(Y)\right] \approx& \mathbb{E}\left[\frac{1}{2}Y'\frac{\partial^2 v(0)}{\partial y \partial y'} Y\right]  \\ 
		= & \frac{1}{2}\mathbb{E}\left[\Tr\left(Y'\frac{\partial^2 v(0)}{\partial y \partial y'}Y\right)\right]\\	
		= & \frac{1}{2}\mathbb{E}\left[\Tr\left(\frac{\partial^2 v(0)}{\partial y \partial y'}YY'\right)\right] \ \ \text{(By commuting within the trace)} \\ 
	 	= & \frac{1}{2}\Tr \left[\frac{\partial^2 v(0)}{\partial y \partial y'}\mathbb{E}\left(YY'\right)\right]\\ 
	 	= & \frac{1}{2} \Tr\left[\frac{\partial^2 v(0)}{\partial y \partial y'}\mathbb{V}(Y)\right]. \\ 
	\end{split}
\end{equation*}

Since $Y$ is assumed close to zero, the variance-covariance matrix is seen as a local risk measure for small risks and by construction does not account for asymmetric risks (which would require an expansion up to order 3), or to extreme risks (which would require an expansion up to order 4). This second--order expansion has been the basis for defining the local version of absolute risk aversion\footnote{In the multidimensional framework, the risk aversion is a matrix directly linked to the Hessian at 0 of function $v$, that is, $\frac{\partial^2v(0)}{\partial y \partial y'}$.} [Arrow (1965)], or for justifying the mean-variance portfolio management in finance [Markowitz (1952), (2000)]\footnote{In the standard financial application with $Y$ as a vector of asset returns, the interest is in the risk of the portfolio returns $\alpha'Y$, where $\alpha$ is the vector of portfolio allocations in values. Then the risk becomes scalar, that is we have $\upsilon(Y)=u(\alpha'Y)$, where $u$ is defined on $\mathbb{R}$. We deduce that the matrix of risk aversion $\frac{\partial^2\upsilon(0)}{\partial y \partial y'}=\frac{d^2u(0)}{dw^2}\alpha \alpha'$ is of rank 1. It involves the effect of portfolio allocation and a scalar risk aversion.}. Since the risk on a price evolution $y=p_{t+h}-p_t$ increases generally with the term, the mean-variance management is appropriate in the short run, that is with frequent predictive updating, and frequent learning. 

\section{Forecast Relative Error Decomposition (FRED)}

As mentioned in section 2.3, the standard FEVD has two drawbacks: (i) While it is appropriate for a local analysis of risk, it is not appropriate for asymmetric risks, or extreme risks. (ii) Moreover it is usually applied to pointwise predictions of $Y$, not to predictions of nonlinear transformations of $Y$. The aim of this section is to provide decompositions that are more appropriate for prediction and learning in a nonlinear dynamic framework. We first provide a general FRED decomposition. This decomposition is applied to one dimensional positive transformations of the process, that are either transition densities, or conditional Laplace transforms of process $(Y_t)$, leading to the so-called Forecast Error Kullback Decomposition (FEKD) and Forecast Error Laplace Decomposition (FELD), respectively. 

\subsection{Relative Forecast Updating}

Let us consider a univariate positive process $(Z_t)$ and the sequence $(I_t)$ of increasing information sets such that $Z_t$ is measurable with respect to $I_t$. We can construct an analogue of the FEVD by considering the relative forecast update:
\begin{equation*}
	\frac{\mathbb{E}(Z_{t+h}|I_{t+h-1})}{\mathbb{E}(Z_{t+h}|I_{t+h})}, \ k=0,...,h-1. 
\end{equation*}
Then we have:
\begin{equation*}
	\frac{Z_{t+h}}{\mathbb{E}(Z_{t+h}|I_t)} = \prod_{k=0}^{h-1}\left[\frac{\mathbb{E}(Z_{t+h}|I_{t+k+1})}{\mathbb{E}(Z_{t+h}|I_{t+k})}\right],
\end{equation*}
By taking the log of both sides and then the conditional expectation given $I_t$, we get the Forecast Relative Error Decomposition (FRED): 
\begin{lemma}[Forecast Relative Error Decomposition (FRED)]
	\begin{equation}\label{fred}
		\begin{split}
		 \mathbb{E}\left\{\log\left[\frac{\mathbb{E}(Z_{t+h}|I_t)}{Z_{t+h}}\right]\bigg\vert I_t\right\} = & \sum_{k=0}^{h-1}\mathbb{E}\left\{\log\left[\frac{\mathbb{E}(Z_{t+h}|I_{t+k})}{\mathbb{E}(Z_{t+h}|I_{t+k+1})}\right]\bigg\vert I_t\right\},\\
		\end{split}
	\end{equation}
where each term in the decomposition is nonnegative. 
\end{lemma}

\textbf{Proof:} The nonegativity is a consequence of Jensen's inequality. For instance, let us consider the left hand side. Then: 
\begin{equation*}
\begin{split}
	& \mathbb{E}\left\{\log\left[\frac{\mathbb{E}(Z_{t+h}|I_t)}{Z_{t+h}}\right]\bigg\vert I_t\right\}  \\ 
	= & \mathbb{E}\left\{-\log\left[\frac{Z_{t+h}}{\mathbb{E}(Z_{t+h}|I_t)}\right]\bigg\vert I_t\right\} \\
	\geq & -\log  \mathbb{E}\left\{\left[\frac{Z_{t+h}}{\mathbb{E}(Z_{t+h}|I_t)}\right]\bigg\vert I_t\right\} \\ 
	&\text{(By Jensen's Inequality applied to the convex function $-\log(\cdot)$)} \\ 
	= & - \log 1 = 0. \\ 
\end{split}
\end{equation*} 
The proof is similar for the terms in the right hand side, noting that $\mathbb{E}(\cdot | I_t) = \mathbb{E}(\mathbb{E}(\cdot |I_{t+h})|I_t)$ by the Law of Iterated Expectation. \qed \\

The FRED is locally a variance decomposition written on the relative error forecasts. Indeed we have locally:
\begin{equation*}
	\begin{split}
		-\log\left[\frac{Z_{t+h}}{\mathbb{E}(Z_{t+h}|I_t)}\right]&=-\log\left\{1+\frac{Z_{t+h}-\mathbb{E}(Z_{t+h}|I_t)}{\mathbb{E}(Z_{t+h}|I_t)}\right\}, \\
	\end{split}
\end{equation*}
and 
\begin{equation*}
\begin{split}
	\mathbb{E}\left\{-\log\left[\frac{Z_{t+h}}{\mathbb{E}(Z_{t+h}|I_t)}\right]\bigg\vert I_t\right\} & \approx \frac{1}{2} \mathbb{E}\left\{\left[\frac{Z_{t+h}-\mathbb{E}(Z_{t+h}|I_t)}{\mathbb{E}(Z_{t+h}|I_t)}\right]^2\bigg \vert I_t\right\}  \\
	& = \frac{1}{2}\mathbb{V}\left[\frac{Z_{t+h}-\mathbb{E}(Z_{t+h}|I_t)}{\mathbb{E}(Z_{t+h}|I_t)}\bigg \vert I_t\right],\\ 
 \end{split}
\end{equation*}
since $\mathbb{E}\left[\frac{Z_{t+h}-\mathbb{E}(Z_{t+h}|I_t)}{\mathbb{E}(Z_{t+h}|I_t)}\bigg\vert I_t\right]=0$. Therefore, the main difference between the FRED and the FEVD is that locally the variance expansion is written on the relative forecast errors instead of the absolute forecast errors and can be applied to any positive transformation $Z_t$ of the multivariate process $(Y_t)$. The FRED \eqref{fred} can be written as:
\begin{equation*}
\gamma(h|I_t) = \sum_{k=0}^{h-1}\gamma(k,h|I_t),
\end{equation*}
where the left hand side of equation \eqref{fred} $\gamma(h|I_t)$, say, measures the risk on the prediction errors at horizon $h$, and the generic term of the right hand side, $\gamma(k,h|I_t)$, has a forward interpretation. More precisely, we have:
\begin{equation*}
	\begin{split}
		\gamma(k,h|I_t) = & \mathbb{E}\left\{\log\left[\frac{\mathbb{E}(Z_{t+h}|I_{t+k})}{\mathbb{E}(Z_{t+h}|I_{t+k+1})}\right]\bigg\vert I_t \right\} \\ 
		 = & \mathbb{E}\left[\mathbb{E}\left\{\log \left[\frac{\mathbb{E}(Z_{t+h}|I_{t+k})}{\mathbb{E}(Z_{t+h}|I_{t+k+1})}\right]\bigg \vert I_{t+k}\right\}\bigg \vert I_t\right]. \\
	\end{split}
\end{equation*}
Therefore, $\gamma(k,h|I_t)$ is the expectation at date $t$ of the risk on the short run prediction error of $Z_{t+h}$ at date $t+k$. As usual, this forward interpretation involves three dates: $t$, $t+k$ and $t+h$. 

\subsection{Forecast Error Kullback Decomposition (FEKD)}

To account for cross-sectional nonlinearities and avoid pointwise predictions, a first solution is to consider the transition predictive densities. Let us assume a strictly stationary Markov process $(Y_t)$ with transition density at horizon $h$ and time $t$ given by $f(y,h|Y_t)$. These densities are positive transforms of $Y_t$. They are well defined for horizons $h\geq 1$, but not at horizon $h=0$, where the value $Y_t$ is perfectly known and the predictive distribution degenerates into a point mass at $Y_t$. Up to this degeneracy, we deduce from Lemma 4 the Forecast Error Kullback Decomposition below:
\begin{proposition}[Forecast Error Kullback Decomposition (FEKD)]
	For any value $y$, we have:
	\begin{equation}\label{FEKD}
	\begin{split}
	 \mathbb{E}\left\{\log\left[\frac{f(y,h|I_t)}{f(y,1|I_{t+h-1})}\right]\bigg\vert I_t\right\} & = \sum_{k=0}^{h-2}\mathbb{E}\left\{\log\left[\frac{f(y,h-k|I_{t+k})}{f(y,h-k-1|I_{t+k+1})}\right]\bigg\vert I_t\right\}.\\
\end{split}
	\end{equation}
\end{proposition}
We get a decomposition that involves measures of risk on the updating of predictive densities. When $Y$ is a continuous variable, the FEKD is invariant by one-to-one differentiable transformations of $Y$. Indeed, due to the ratios, the Jacobian effect disappears in the decomposition formula. The generic term on the right hand side of the FEKD is: 
\begin{equation*}
\gamma(k,h|y,I_t) =\mathbb{E}\left\{\log\left[\frac{f(y,h-k|I_{t+k})}{f(y,h-k-1|I_{t+k+1})}\bigg\vert I_{t+k}\right]\bigg\vert I_{t}\right\},
\end{equation*}
that is the conditional Kullback proximity measure (or contrast, or divergence) between the two conditional densities. \\

A main difference between the FEKD and FEVD is that the decomposition \eqref{FEKD} can be written for any value of $y$, whereas there is a single FEVD. Typically, let us consider a univariate stationary process $(Y_t)$. From $T$ observations $Y_1,...,Y_T$, we can derive the sample distributions, then the sample deciles, and compare the FEKD evaluate at difference deciles. This will provide a more detailed analysis of risk with clear applications to the dynamic analysis of inequality, or the analysis of financial risks (where these quantiles are usually called Value-at-Risk (VaR)). \\

When the process $(Y_t)$ has a dimension larger than 2, the notion of quantile has not yet been defined, and it is less clear how to select an appropriate grid of values $y$ to which the FEKD will be applied. A solution can be to apply the approach by deciles to combinations of components of $Y_t$, if such combinations have an economic meaning. Combinations are more easily treated by means of Laplace transforms, as seen in the next subsection. Note that the FEKD can exist in cases where the FEVD has no meaning. Indeed, the FEVD requires that the observed $Y_t$ are square integrable, so it cannot be applied to data with fat tails. We give below in Section 4.1.2 the example of the Cauchy AR(1) model, where the FEKD exists, but not the FEVD. 

\subsection{Forecast Error Laplace Decomposition (FELD)}

Let us consider a process $(Y_t)$ of dimension $n$. Its conditional Laplace transform at horizon $h$ is defined by:
\begin{equation}
	\Psi(u,h|I_t) = \mathbb{E}\left[\exp(-u'Y_{t+h})|I_t\right],
\end{equation}
where $u\in D \subset \mathbb{R}^n$ and such that the expectation exists on domain $D$. It is known that the knowledge of the Laplace transform is equivalent to the knowledge of the conditional distribution for the Gaussian case, or if the process satisfies some positivity restrictions [see Feller (1971) and the examples in Section 4]. Then we can apply the FRED to these nonlinear transformations.
\begin{proposition}[FELD]
\begin{equation}\label{FELD}
	\begin{split}
	\mathbb{E}\left\{\log\left[\frac{\Psi(u,h|I_t)}{\exp(-u'Y_{t+h})}\right]\bigg\vert I_t\right\} & = \sum_{k=0}^{h-1}\mathbb{E}\left\{\log\left[\frac{\Psi(u,h-k|I_{t+k})}{\Psi(u,h-k-1|I_{t+k+1})}\right]\bigg\vert I_t\right\},\\
\end{split}
\end{equation}
for any $u\in D$, where $D$ is the domain of arguments in ensuring the existence of the conditional Laplace transforms. 
\end{proposition}

As for the FEKD, the FELD \eqref{FELD} defines several decompositions: $\gamma(h|u,I_t)=\sum_{k=0}^{h-1}\gamma(k,h|u,I_t)$, as many as selected combinations $u$, $u\in D$. In some applications such combinations can have economic interpretations to define portfolio allocations when $Y_t$ is a vector of $n$ asset prices at date $t$, or to combine income and wealth in the analysis of inequality. When $(Y_t)$ satisfies some ``positivity" restrictions, the Laplace transform with positive argument $u$ characterizes the distribution. Moreover, we have $\mathbb{E}\left[\exp(-u'Y_{t+h})\vert I_t\right]\leq 1$, and the FELD always exists, even if the distribution of $Y_t$ has very fat tails. 

\subsection{Decomposition of Risk Premium}

The FELD can be interpreted as a decomposition of risk premiums for spot prices. Consider a decision maker with exponential utility function $u(y)=-\exp(-uy)$. The certainty equivalent $\pi(u)$ is the level of wealth which makes the decision maker indifferent from the expected outcome of a lottery on $Y$. That is: 
\begin{equation*}
	\begin{split}
			\pi(u)  & = -\frac{1}{u}\log\mathbb{E}\left[\exp(-uY)\right]\equiv -\frac{1}{u}\log\Psi(u),\\
	\end{split}
\end{equation*}
which is a function of parameter $u$, which is the Arrow-Pratt scalar measure of risk aversion. Equivalently, for a risky asset with price $Y_t > 0$, we can write: 
\begin{equation}
	\pi(u,h|I_t) = -\frac{1}{u} \log \Psi(u,h|I_t),
\end{equation}
which is the spot value of the asset at time $t$ and horizon $h$. Intuitively, it is a contract written at time $t$, which captures the value of delivering the asset at some future horizon $h$.\\


For an investor with risk aversion parameter $u$, the FELD in \eqref{FELD} can be written as:
\begin{equation}\label{FELDp}
	\begin{split}
		\mathbb{E}\left[-\frac{1}{u}\log \Psi(u,h|I_t)+Y_{t+h}\bigg \vert I_t\right] & = \sum_{k=0}^{h-1}\mathbb{E}\left\{-\frac{1}{u}\log\Psi(u,h-k|I_{t+k}) + \frac{1}{u} \log \Psi(u,h-k-1|I_{t+k+1})\bigg\vert I_t\right\}\\
		\iff \mathbb{E}[Y_{t+h}-\pi(u,h|I_t)|I_t]& = \sum^{h-1}_{k=0} \mathbb{E}\left[\pi(u,h-k|I_{t+k})-\pi(u,h-k-1|I_{t+k+1})\bigg \vert I_t\right] \\ 
		\iff \pi(u,h|I_t) - \mathbb{E}(Y_{t+h}|I_t)&= \sum^{h-1}_{k=0} \mathbb{E}\left[\pi(u,h-k-1|I_{t+k+1})-\pi(u,h-k|I_{t+k})\bigg \vert I_t\right]. \\
	\end{split}
\end{equation}
Let us now consider the economic interpretations of this decomposition formula. The term $\pi(u,h|I_t) - \mathbb{E}(Y_{t+h}|I_t)$ is the difference between the value (price) of $Y_{t+h}$ at date $t$ and its historical conditional expectation. This difference is positive (by Jensen's inequality) and usually interpreted as a risk premium. Therefore, \eqref{FELDp} provides a decomposition of this risk premium. More precisely, the term $\pi(u,k,h|I_t)=\mathbb{E}\left[\pi(u,h-k|I_{t+k})\bigg \vert I_t \right]$ is the value of a forward contract of the asset, written at time $t$, for a payment at time $t+k$ and delivery at time $t+h$. Hence, the generic term on the RHS of \eqref{FELDp} captures the difference in values $\pi_f$ of the forward contracts for payment at time $t+k$ and $t+k+1$ for the delivery of the asset at time $t+h$. As $h$ varies, we get a decomposition of the term structure for the risk premium, or equivalently of the spot value (price) as: 
\begin{equation}
\pi(u,h|I_t) = \mathbb{E}(Y_{t+h}|I_t)+ \sum^{h-1}_{k=0} \mathbb{E}\left[\pi_f(u,k+1,h|I_{t+k+1})-\pi_f(u,k,h|I_{t+k})\bigg \vert I_t\right],
\end{equation}
that is compatible with the no dynamic arbitrage condition between spot and forward contracts. There is a debate on the valuation approach to be chosen for contingent assets [see e.g. Embrechts (2000)]. For financial assets traded on very liquid markets, this is usually done by introducing a stochastic discount factor to satisfy the no dynamic arbitrage opportunity condition. The situation is different for individual insurance contracts or for operational risks [see the discussion of cyber risk in Section 7]. The certainty equivalent principle is a more appropriate valuation approach in such frameworks and we have checked ex-post that it is compatible with the no arbitrage opportunity assumption. 

\section{Examples}

In this section, we consider different dynamic models for which we derive closed form decompositions for either the FEKD, or the FELD. 

\subsection{Examples of FEKD}

\subsubsection{The Gaussian VAR(1)}

Let us assume that the $n$-dimensional stationary process $(Y_t)$ satisfies:
\begin{equation}\label{VAR}
	Y_t = \Phi Y_{t-1} + \varepsilon_t,
\end{equation}
where the eigenvalues of $\Phi$ have a modulus strictly smaller than 1 and the $\varepsilon_t$'s are i.i.d. Gaussian $\varepsilon_t \sim N(0,\Sigma)$. Then, the conditional distribution of $Y_{t+h}$ given $Y_t$ is Gaussian with mean $\Phi^hY_t$ and variance-covariance matrix $\Sigma_h = \Sigma + \Phi \Sigma \Phi'+...+\Phi^{h-1}\Sigma(\Phi')^{h-1}$. The following proposition provides the closed form FEKD for the Gaussian VAR(1). 
\begin{proposition}
	In the Gaussian VAR(1) model, the FEKD is of the form:
	\begin{equation*}\label{FEKD_AR}
		\begin{split}
			&	a(h|Y_t) + b(h|Y_t)y+y'c(h|Y_t)y \\
			& = \sum_{k=0}^{h-2} \left[a(h,k|Y_t) + b(h,k|Y_t)y+y'c(k,h|Y_t)y\right],\\ 
		\end{split}
	\end{equation*}
	where:
	\begin{equation*}
		\begin{split}
		a(h,k|Y_t) =  & \frac{1}{2}\log\left[\frac{\det\Sigma_{h-k-1}}{\det\Sigma_{h-k}}\right] + \frac{1}{2}\Tr \left[\Sigma^{-1}_{h-k-1}\Phi^{h-k-1}\Sigma_{k+1}\left(\Phi^{h-k-1}\right)'-\Sigma^{-1}_{h-k}\Phi^{h-k}\Sigma_k\left(\Phi^{h-k}\right)'\right]  \\
		& -\frac{1}{2}Y_t'\left(\Phi^h\right)'\left(\Sigma^{-1}_{h-k}-\Sigma^{-1}_{h-k-1}\right)\Phi^hY_t,\\
		b(h,k|Y_t) & =  Y_t'\left(\Phi^h\right)'\left(\Sigma^{-1}_{h-k}-\Sigma^{-1}_{h-k-1}\right), \\ 
		c(h,k|Y_t) & = - \frac{1}{2}\left(\Sigma^{-1}_{h-k}-\Sigma^{-1}_{h-k-1}\right). \\ 
		\end{split}
	\end{equation*}
\end{proposition}
\textbf{Proof:} See the Appendix A.1.1. \\ 

For the Gaussian VAR(1) the FEVD is:
\begin{equation}
	\Sigma_h - \Sigma = \sum_{k=0}^{h-2} \left[\Sigma_{h-k} - \Sigma_{h-k-1} \right].
\end{equation}
The FEKD in Proposition 3 differs from the FEVD, since it depends on $Y_t$ and on the argument $y$. Its functional quadratic form in $y$ provides information on the prediction errors and their decomposition in the tails when we focus on the components in $y$ and the squares. The decomposition of the intercept is just to balance the change in tails, since all the predictive distributions are unit mass. The component in $y$ depends on $Y_t$, which corresponds to the forward interpretation of the elements in the decomposition. The quadratic element is independent of $Y_t$ due to the conditional homosecedasticity of the Gaussian VAR(1) model. This element is written on the inverses of the prediction variance, not on the prediction variances as in the FEVD. This difference is the analogue of the two equivalent filters, introduced by Kalman for the linear state space model. Indeed, the standard covariance filter is based on the direct updating of $\Sigma_h$, whereas the information form of the filter coincides with the direct updating of the inverse $\Sigma^{-1}_h$ (called information). 

\subsubsection{The Cauchy AR(1)}

Let us consider the stationary univariate process $(Y_t)$ defined by:
\begin{equation*}
	Y_t = \varphi Y_{t-1} + \sigma\varepsilon_t, \ \ |\varphi|<1,
\end{equation*}
where $(\varepsilon_t)$ is a Cauchy distributed strong white noise. This distribution admits the density $f(\varepsilon)=\frac{1}{\pi}\frac{1}{1+\varepsilon^2}$, and its characteristic function is given by: $\mathbb{E}(\exp(iu\varepsilon))= \exp(-|u|)$, where $i$ is the imaginary number. Then we have:
\begin{equation*}
	\begin{split}
		Y_{t+h} = \varphi^h Y_t + \sigma(\varepsilon_{t+h}+\varphi\varepsilon_{t+h-1}+...+\varphi^{h-1}\varepsilon_{t+1}) \equiv \varphi^hY_t + \varepsilon_{t,h},
	\end{split}
\end{equation*}
where:
\begin{equation*}
	\begin{split}
& \mathbb{E}\left[\exp(iu\varepsilon_{t,h})\right] \\
= & \mathbb{E}\left\{\exp\left[iu(\sigma\varepsilon_{t+h}+...+\sigma\varphi^{h-1}\varepsilon_{t+1})\right]\right\} \\ 
= & \exp\left[-|u|(\sigma+\sigma|\varphi|+...+\sigma|\varphi|^{h-1})\right]\\
= & \exp\left[-|u|\sigma \frac{1-|\varphi|^h}{1-|\varphi|}\right].\\ 
	\end{split}
\end{equation*} 
Therefore the conditional distribution of $Y_{t+h}$ given $Y_t$ is a Cauchy distribution with drift $\varphi^hY_t$ and scale $\sigma \frac{1-|\varphi|^h}{1-|\varphi|}$. We deduce that:
\begin{equation*}
	\frac{f(y,h-k|I_{t+k})}{f(y,h-k-1|I_{t+k+1})}=\frac{1-|\varphi|^{h-k-1}}{1-|\varphi|^{h-k}}\frac{1+\left[(y-\varphi^{h-k-1}Y_{t+k+1})/\left(\sigma\frac{1-|\varphi|^{h-k-1}}{1-|\varphi|}\right)\right]^2}{1+\left[(y-\varphi^{h-k}Y_{t+k}) /\left(\sigma\frac{1-|\varphi|^{h-k}}{1-|\varphi|}\right)\right]^2},
\end{equation*}
and then:
\begin{equation*}
	\begin{split}
	\gamma(k,h|y,I_t) = & \log\left[\frac{1-|\varphi|^{h-k-1}}{1-|\varphi|^{h-k}}\right]\\
	&+\mathbb{E}\left\{\log\left[1+\left((y-\varphi^{h-k-1}Y_{t+k+1})\bigg/\left(\sigma\frac{1-|\varphi|^{h-k-1}}{1-|\varphi|}\right)\right)^2\right]\bigg \vert I_t\right\}\\
	&-\mathbb{E}\left\{\log\left[1+\left((y-\varphi^{h-k}Y_{t+k})\bigg/\left(\sigma\frac{1-|\varphi|^{h-k}}{1-|\varphi|}\right)\right)^2\right]\bigg \vert I_t\right\}.\\
	\end{split}
\end{equation*}
This is the generic term of the RHS of the FEKD. In this framework the conditional variance at horizon $h$, that is $\mathbb{V}\left(Y_{t+h}|Y_t\right)$, does not exist and the FEVD does not exist as well. However, the elements $\gamma(k,h|I_t)$ in the decomposition above exist, since the transformed variable $\log(\alpha+\beta\varepsilon_t^2)$ is integrable with respect to the Cauchy distribution. 
\subsubsection{Markov Chain}

Let us start by considering a stationary Markov chain with two states $0,1$, or equivalently, a stationary Markov binary time series $(Y_t)$. The transition matrix of the chain can be parameterized by the marginal probability $\pi = P(Y_t =1)$ and a persistence parameter $\lambda$, $0 \leq \lambda < 1$. We have:
\begin{equation*}
	P(Y_{t+1}=1|Y_t) = \mathbb{E}(Y_{t+1}|Y_t) = \pi + \lambda (Y_t-\pi), \ Y_t=0,1.
\end{equation*}
Then by iterated expectations we deduce:
\begin{equation*}
	P(Y_{t+h}=1|1,Y_t) = \mathbb{E}(Y_{t+h}|Y_t) = \pi + \lambda^h(Y_t-\pi), \ Y_t=0,1.
\end{equation*}
\begin{proposition}
	For a binary Markov chain with states 0 and 1, the generic term of the right hand side of the FEKD \eqref{FEKD} for $y=1$ is: 
	\begin{equation}
		\begin{split}\label{fekd_mc_bin}
			\gamma(k,h|Y_t)= & \log\left[\frac{1-\lambda^{h-k}}{1-\lambda^{h-k-1}}\right] + \log\left[\frac{\pi+\lambda^{h-k}(1-\pi)}{\pi(1-\lambda^{h-k})}\right]\left[\pi+\lambda^k(Y_t-\pi)\right] \\
			& -  \log\left[\frac{\pi+\lambda^{h-k-1}(1-\pi)}{\pi(1-\lambda^{h-k-1})}\right]\left[\pi+\lambda^{k+1}(Y_t-\pi)\right], \\
		\end{split}
	\end{equation}
which is in the form of $\alpha(h,k)Y_t + \beta(h,k)$.
\end{proposition}
\textbf{Proof:} See Appendix A.1.2. \\ 

Such univariate binary processes have attracted considerable attention to describe the business cycle recession/expansion periods [see e.g. Estrella, Mishkin (1998), Chauvet, Potter (2005), or Kauppi, Saikkonen (2008)]. \\

This term can be compared with the generic term in the standard FEVD.
\begin{corollary}
	The FEVD for the binary Markov chain is given by:
	\begin{equation*}
\begin{split}
			& \mathbb{V}\left\{\mathbb{E}\left[Y_{t+h}|I_{t+k+1}\right]-\mathbb{E}\left[Y_{t+h}|I_{t+k}\right]\bigg \vert I_t \right\} \\
= & \pi(1-\pi)\lambda^{2(h-k-1)}\left[1-\lambda^2\right]- (Y_t-\pi)\lambda^{2h-k-1}\left[1-2\pi\right]\left[1-\lambda\right]. \\ 
\end{split}
	\end{equation*}
\end{corollary}
\textbf{Proof:} See Appendix A.1.3. \\ 

Now we can deduce the extension to the general case. Suppose $Y_t$ is a stationary Markov chain with $n$ possible states. Let $X_t = (X_{1,t},...,X_{n,t})'$, with $X_{i,t}=1$, if $Y_t$ is in state $i$, and zero otherwise, for $i=1,...,n$. The knowledge of $Y_t$ is equivalent to the knowledge of $X_t$, which can take on the values: $(1,0,...0)'$, $(0,1,0,...,0),...,(0,...,0,1)'$. Let us characterize its $h$-step transition probabilities with the matrix $P^{h}$, whose elements are defined as $p^{(h)}_{ij} = P(Y_{t+h}=i|Y_t=j)$ for all $t\geq0$. Then, the FEKD is given by the following proposition. 
\begin{proposition}
	For a Markov chain with $n$ states, the generic term on the right hand side of the FEKD is:
	\begin{equation*}
		\begin{split}
	\gamma(h,k|I_{t+k}) & = \left[\widetilde{\log}(P^{h-k})_yP^{k+1}-\widetilde{\log}(P^{h-k-1})_yP^{k}\right]X_t,
		\end{split}
	\end{equation*}	
where $\widetilde{\log}(P^{h})$ is a matrix whose elements are the logged elements of $P^{h}$ and $A_y$ denotes the y-th row of matrix $A$.
\end{proposition}
\textbf{Proof:} See Appendix A.1.4. \\

The result presented in Propostion 4 is a special case of Proposition 5. 

\begin{corollary}
	The expression of the FEKD for the binary Markov chain in \eqref{fekd_mc_bin} is a special case of Proposition 5 where: 
\begin{equation*}
	P^{h}= \begin{bmatrix}
		p^{(h)}_{00}&p^{(h)}_{01} \\
		p^{(h)}_{10}&p^{(h)}_{11} \\ 
	\end{bmatrix}= \begin{bmatrix}
		1-\left[\pi(1-\lambda^{h})\right] &1-\left[\pi+\lambda^{h}(1-\pi)\right]  \\
		\pi(1-\lambda^{h})& \pi+\lambda^{h}(1-\pi) \\
	\end{bmatrix}.
\end{equation*}
\end{corollary}

\textbf{Proof:} See Appendix A.1.5. \\ 

\subsection{Examples of FELD}

The FELD has a simple form and is easy to interpret for the dynamic affine model, which is called the Compound Autoregressive (CaR) model in discrete time [see Duffie, Filipovic, Schachermayer (2003), Darolles, Gourieroux, Jasiak (2006)]. We first review the CaR models and some of their dynamic properties. Then, we discuss in detail the application to Gaussian processes, Integer Autoregressive (INAR) models, Negative Binomial Autoregressive (NBAR) models, Markov Chains, Autoregressive Gamma  and Wishart processes. Remember that the FELD has an interpretation as a decomposotion of risk premium where $Y$ is a value (see Section 3.4).

\subsubsection{Dynamic Affine Model}
The process is assumed Markov of order 1 with a conditional log-Laplace transform which is affine in the conditioning value\footnote{The extension to a Markov of order $p$ is straightforward.} $Y_t$. Thus, the Laplace transform at horizon 1 can be written as:
\begin{equation}
		\begin{split}
		\Psi(u,1|I_t) = & \Phi(u,1|Y_t) \\
		=& \mathbb{E}\left[\exp(-u'Y_{t+1})|Y_t\right] \\ 
		 = & \exp \left\{-a(u)'Y_t + c(u)-c[a(u)]\right\}, \\
	\end{split}
\end{equation}
where $c(u) =\log \mathbb{E}[\exp(-u'Y_t)]$ is the unconditional log-Laplace transform of $Y_t$ and the function $a(\cdot)$ captures all nonlinear serial dependence features. The affine property remains satisfied at any forecast horizon. More precisely, we have: 
\begin{equation}\label{cllt}
	\Psi(u,h|Y_t) = \exp\left\{-a^{\circ h}(u)' Y_t + c(u) - c\left[a^{\circ h}(u)\right]\right\},
\end{equation}
where $a^{\circ h}(\cdot)$ is function $a(\cdot)$ compounded $h$ times with itself. Then, the FELD becomes:
\begin{equation}
		\begin{split}
		& \mathbb{E}\left[u'Y_{t+h}-a^{\circ h}(u)'Y_t+c(u)-c[a^{\circ h}(u)]\bigg \vert I_t \right] \\
	= & \sum_{k=0}^{h-1}\mathbb{E}\left\{a^{\circ (h-k-1)}(u)Y_{t+k+1}-a^{\circ (h-k)}(u)'Y_{t+k}+c\left[a^{\circ (h-k-1)}(u)\right]-c\left[a^{\circ (h-k)}(u)\right]\bigg \vert I_t \right\}, \ \forall u.  \\ 
	\end{split}
\end{equation}
This decomposition involves the conditional expectations of $Y_{t+k}$ given $Y_t$ and can be written as:
\begin{equation}
	\begin{split}
		& u'\mathbb{E}(Y_{t+h}|Y_t) - a^{\circ h}(u)'Y_t + c(u) -c\left[a^{\circ h}(u)\right] \\
		= & \sum_{k=0}^{h=1} \left\{a^{\circ(h-k-1)'}(u)\mathbb{E}(Y_{t+k+1}|Y_t)-a^{\circ(h-k)'}(u)\mathbb{E}(Y_{t+k}|Y_t)\right\} + c\left[a^{\circ(h-k-1)}(u)\right]- c\left[a^{\circ(h-k)}(u)\right], \ \forall u.\\ 
	\end{split}
\end{equation}
Whereas the FELD involves the conditional expectations only, the nonlinear dynamic features are taken into account by the ``weightings" of the expectation that depend on the function $a(\cdot)$, that summarizes the nonlinear serial dependence. \\

It is known that the log-Laplace transform admits a Taylor expansion in terms of cumulants in a neighborhood of $u=0$. In particular, the first-order expansion of the conditional log-Laplace transform shows that the conditional expectation is affine in the conditioning variable $Y_t$. More precisely, we have:
\begin{equation*}
\mathbb{E}(Y_{t+1}|Y_t) = -\frac{dc}{du}(0) + \frac{da'}{du}(0) \left[Y_t + \frac{dc}{du}(0)\right].
\end{equation*}
and
\begin{equation*}
\mathbb{E}(Y_{t+h}|Y_t) = -\frac{dc}{du}(0) +\left[\frac{da'}{du}(0)\right]^h\left[Y_t + \frac{dc}{du}(0)\right].
\end{equation*}
Then we deduce a closed form FELD for dynamic affine models. 
\begin{proposition}
For dynamic affine models, the FELD takes the form:
\begin{equation*}
\begin{split}
	& \left\{u'\left[\frac{da'}{du}(0)\right]^h - a^{\circ h}(u)'\right\}Y_t - u'\frac{dc}{du}(0)+ u'\left[\frac{da'}{du}(0)\right]^h\left[\frac{dc}{du}(0)\right] + c(u) - c\left[a^{\circ h}(u)\right] \\ 
	= & \sum_{k=0}^{h-1}\left\{a^{\circ (h-k-1)}(u)'\left[\frac{da'}{du}(0)\right]^{k+1}-a^{\circ (h-k)}(u)'\left[\frac{da'}{du}(0)\right]^{k}\right\}Y_t+\left(a^{\circ (h-k)}(u)'-a^{\circ (h-k-1)}(u)'\right)\left[\frac{dc}{du}(0)\right]\\
	& + \left(a^{\circ (h-k)}(u)'\left[\frac{da'}{du}(0)\right]^k-a^{\circ (h-k-1)}(u)'\left[\frac{da'}{du}(0)\right]^{k+1}\right)\left[\frac{dc}{du}(0)\right] + c\left[a^{\circ (h-k-1)}(u)\right]-c\left[a^{\circ (h-k)}(u)\right], \ \forall u.\\  
	\end{split}
\end{equation*}
\end{proposition}

\textbf{Proof:} See Appendix A.2.1. \\ 

Therefore, we get a decomposition of the type:
\begin{equation}
	\alpha(h,u)'Y_t + \beta(h,u) = \sum_{k=0}^{h-1} \left[\alpha(h,k,u)'Y_t + \beta(h,k,u)\right],
\end{equation}
or equivalently decompositions of the functions $\alpha(h,u)$ and $\beta(h,u)$ into $\sum_{k=0}^{h-1}\alpha(h,k,u)$ and $\sum_{k=0}^{h-1}\beta(h,k,u) $ respectively. These decompositions depend on both the term $h$ and the argument $u$. The decomposition of function $\alpha(h,u)$ is especially appealing due to its interpretation in terms of nonlinear Impulse Response Functions (IRF). Indeed, let us consider a given shock of magnitude $\delta$ on the level $Y_t$. Note that in our nonlinear framework, the magnitude of the shock $\delta$ is constrained by the domain of $Y_t$. It can be multivariate if $Y_t$ is multivariate, constrained to be either 0, or -1 (resp. 0, or 1) if $Y_t$ is binary with value 1 (resp. value 0), and so on. We do not discuss the sources of the shock and if they are identifiable or controllable. The effect on the predictive distribution at horizon $h$ is $\alpha(h,u)'\delta$. Compared to the standard linear approach of the IRF, we see that the IRF depends on the argument $u$. In other words, this measure changes with the preference (risk aversion) of the analyst. Moreover, for a given $u$, the decomposition will change, since in a nonlinear dynamic forecast, the spot and forward short run updatings of the predictive distributions will differ. \\

The FELD can be easily compared to the FEVD. Let us consider the one-dimensional case for expository purposes. In an affine dynamic model, the conditional variance $\mathbb{V}(Y_{t+h}|I_{t+k})$ is an affine function of $Y_{t+k}$, and thus the generic term in the FEVD \eqref{FEVD} is also an affine function of $Y_t$. More precisely, we get:
\begin{corollary}
	\begin{equation*}
		\begin{split}
 & \mathbb{E}\left\{\mathbb{V}\left[\mathbb{E}\left(Y_{t+h}\vert I_{t+k+1}\right)\vert I_{t+k}\right]\vert I_t\right\}  \\ 
 = & \sum_{j=1}^N \left\{\left(\frac{\partial a'(0)}{du}\right)^{h-k-1}\frac{\partial^2a_j(0)}{\partial u \partial u'}\left(\frac{\partial a(0)}{du'}\right)^{h-k-1}\left[-\frac{dc(0)}{du_j}+\left(\left(\frac{\partial a'(0)}{\partial u}\right)^k\left(Y_t+\frac{dc(0)}{du}\right)\right)_j\right]\right\}\\ 
 & + \left(\frac{\partial a'(0)}{\partial u}\right)^{h-k-1}\frac{\partial^2b(0)}{\partial u \partial u'}\left(\frac{\partial a(0)}{\partial u'}\right)^{h-k-1}.\\
		\end{split}
	\end{equation*}
\end{corollary}
\textbf{Proof:} See Appendix A.2.2.

\subsubsection{Strong Linear VAR(1) Model}
Let us consider the strong VAR(1) model:
\begin{equation}
	Y_t=\Phi Y_{t-1} +\varepsilon_t,
\end{equation}
where the $\varepsilon_t$'s are i.i.d with the log-Laplace transform:
\begin{equation}
	\log \mathbb{E}\left[\exp(-u'\varepsilon_{t+1})\right] = b(u),
\end{equation}
The conditional Laplace transform is:
\begin{equation}
	\mathbb{E}\left[\exp(-u'\varepsilon_{t+1})|Y_t\right] = \exp \left[-u'\Phi Y_t +b(u)\right].
\end{equation}
When the eigenvalues of $\Phi$ have a modulus strictly smaller than 1, we can write the infinite moving average representation of process $(Y_t)$ as:
\begin{equation}
 Y_t = \sum_{j=0}^\infty \Phi^j \varepsilon_{t-j}.
\end{equation}
Therefore, the unconditional log-Laplace transform of process $(Y_t)$ is:
\begin{equation}
	c(u) = \sum_{j=0}^{\infty} b\left[(\Phi')^ju\right].
\end{equation}
The strong VAR(1) is an affine model, with $a(u)=\Phi'u$ and $a^{\circ h}(u)=(\Phi')^hu$. Since the dynamic is linear and the function $a(\cdot)$ is also linear, the FELD is greatly simplified. We get: 
\begin{corollary}
	For a strong linear VAR(1) model, the FELD is: 
\begin{equation}
 \mathbb{E}\left\{\log\left[\frac{\mathbb{E}\left[\exp\left(-u'Y_{t+h}\right)|I_t\right]}{\exp(-u'Y_{t+h})}\right]\bigg\vert I_t\right\} = \sum_{k=0}^{h-1}b\left[\left(\Phi'\right)^ku\right].
\end{equation}
\end{corollary}

%

Due to the linear dynamic, the FELD does not depend on the value $Y_t$ of the conditioning variable. However, there are still effects on the decomposition of the cross-sectional heterogeneity, that is the non-Gaussian distribution of the errors $(\varepsilon_t)$. If the error is Gaussian $\varepsilon_t \sim IIN(0,\Sigma)$, we get $b(u)=\frac{u'\Sigma u}{2}$. Then, the right hand side of the decomposition becomes:
\begin{equation*}
		\begin{split}
		\sum_{k=0}^{h-1} b\left[\left(\Phi'\right)^k u\right] = & \frac{1}{2}\sum_{k=0}^{h-1} \left(u'\Phi^k \Sigma \left(\Phi'\right)^ku\right)\\ 
		= & \frac{1}{2}u'\sum_{k=0}^{h-1} \Phi^k \Sigma \left(\Phi'\right)^ku \\
	\end{split}
\end{equation*}
We recover the right hand side $\sum_{k=0}^{h-1} \Phi^k \Sigma \left(\Phi'\right)^k$ in the FEVD for the VAR(1) model (see equation (2.8)). To summarize:
\begin{enumerate}
	\item In a strong linear dynamic model, the FELD does not depend on the conditioning variable.
	\item The decomposition is equivalent to the FEVD only if the white noise is Gaussian.
\end{enumerate}
\subsubsection{Markov Chain}

Let us return to the  Markov chain example discussed in section 4.1.2 with $n$ states. The conditional log-Laplace transform is given by:
\begin{equation*}
	\log \Psi(u,h|X_{t}) = \widetilde{\log} \left[\exp(u)P^{h}\right]X_{t},
\end{equation*}
where $u=(u_1,...,u_n)'$ and $\widetilde{\log}(A)$ is a matrix whose elements are the logged elements of matrix $A$. Then the FELD is given by the following proposition:
\begin{corollary}
For a stationary Markov chain $(Y_t)$ with $n$ states and transition matrix $P$, the FELD is of the form: 
\begin{equation*}
	\begin{split}
		\mathbb{E}\left\{\log\left[\frac{\Psi(u,h|I_t)}{\exp(u'Y_{t+h})}\right]\bigg\vert I_t\right\} & = \sum_{k=0}^{h-2}\left[\widetilde{\log}\left(\exp(u)P^{h-k}\right)P^{k}-\widetilde{\log}\left(\exp(u)P^{h-k-1}\right)P^{k+1}\right]X_t,\\
	\end{split}
\end{equation*}
for all $u=(u_1,...,u_n)'$. 
\end{corollary}
\textbf{Proof:} See Appendix A.2.3.

\subsubsection{INAR Model}

For expository purposes, let us consider the Integer Autoregressive model (INAR) of order 1  introduced by McKenzie (1985), Al-Osh, Azaid (1987). The process is defined by:
\begin{equation*}
	Y_t = \mathcal{B}_t(p) \circ Y_{t-1} + \varepsilon_t,
\end{equation*}
with $\mathcal{B}_t(p) \circ Y_{t-1}=\sum_{j=1}^{Y_{t-1}}U_{j,t}$, where the variables $\varepsilon_t$, $U_{j,t}$, $j,t$ varying, are independent, $U_{j,t}$ follows the same Bernoulli distribution $\mathcal{B}(1,p)$ and $\varepsilon_t$ the Poisson distribution $P(\lambda)$, with $0 \leq p < 1$ and $\lambda\geq0$. By convention, $\sum_{j=1}^{0}U_{j,t} = 0$. The INAR model is a CaR model, with $a(u) = -\log\left[p\exp(-u)+1-p\right]$, $c(u)=\frac{-\lambda}{1-p}\left[1-\exp(-u)\right]$. In particular, the marginal distribution of $Y_t$ is Poisson $P\left(\frac{\lambda}{1-p}\right)$. It is easily checked that: $a^{\circ h}(u)=-\log\left[p^h\exp(-u)+1-p^h\right]$. 

\begin{corollary}
	For the INAR(1) model, the FELD is given by:
	\begin{equation}\label{FELD_INAR}
		\begin{split}
		& \mathbb{E}\left\{\log\left[\frac{\mathbb{E}\left[\exp\left(-u'Y_{t+h}\right)|I_t\right]}{\exp(-u'Y_{t+h})}\right]\bigg\vert I_t\right\} \\
		=& \left\{p^hu+\log\left[1-p^h+p^h\exp (-u)\right]\right\}Y_t+\frac{\lambda}{1-p}\left[\left(1-p^h\right)\left(u-1+\exp(-u)\right)\right]\\ 
= & \sum_{k=0}^{h-1}\left\{\left[p^k\log\left(1-p^{h-k}+p^{h-k}\exp (-u)\right)-p^{k+1}\log\left(1-p^{h-k-1}+p^{h-k-1}\exp (-u)\right)\right]\right\}Y_t\\
&\left. +\lambda \frac{(1-p^k)}{1-p}\log \left(1-p^{h-k}+p^{h-k}\exp(-u)\right)-\lambda \frac{(1-p^{k+1})}{1-p}\log \left(1-p^{h-k-1}+p^{h-k-1}\exp(u)\right)\right.\\ 
&\left. +\lambda \left[1-\exp(-u)\right]p^{h-k-1}\right..\\ 
	 \end{split}
	\end{equation}
\end{corollary}
\textbf{Proof:} See Appendix A.2.4. \\

\subsubsection{Cox, Ingersoll, Ross and Autoregressive Gamma Process}

Let us consider the univariate Autoregressive Gamma Process of order 1, which is the time discretized Cox, Ingersoll, Ross model [see Gourieroux, Jasiak (2006)]. This dynamic model is the benchmark for the dynamic analysis of short run interest rates, or for one dimensional stochastic volatility. This is an affine model where the transition distribution is constructed from a gamma distribution with path dependent stochastic degree of freedom. It depends on two parameters: $\beta\geq 0$, $\delta\geq0$, and the Laplace transform corresponding to $a(u)=\frac{\beta u}{1+u}$, $c(u)=-\delta \log\left[1+\frac{u}{1-\beta}\right]$. In particular, the marginal distribution of the process is such that: $(1-\beta)Y_{t} \sim \gamma(\delta)$. It is easily checked that $a^{\circ h}(u)=\frac{\beta^h u}{\left[1+\frac{1-\beta^h}{1-\beta}u\right]}$ and that $\frac{da(u)}{du} = \beta$. Therefore, we get the following FELD decomposition of the functional IRF (the decomposition of the intercept $\beta(h.u)$ is provided in Appendix A.2.5). 

\begin{corollary}
	For the ARG(1) process, the FELD leads to the decomposition of:
	\begin{equation*}
	\alpha(h,u) = u\beta^h\left[1-\frac{1}{1+\frac{1-\beta^h}{1-\beta}u}\right],
	\end{equation*}
	into $\sum_{k=0}^{h-1} \alpha(h,k,u)$ where:
	\begin{equation*}
		\alpha(h,k,u) = u\beta^h \left[\frac{1}{1+\frac{1-\beta^{h-k-1}}{1-\beta}u}-\frac{1}{1+\frac{1-\beta^{h-k}}{1-\beta}u}\right]
	\end{equation*}
\end{corollary}
\textbf{Proof:} See Appendix A.2.5. \\

Hence, the proportion $\frac{\alpha(h,k,u)}{\alpha(h,u)}$ is given by: 
\begin{equation*}
	\frac{\alpha(h,k,u)}{\alpha(h,u)}= \frac{ \left[\frac{1}{1+\frac{1-\beta^{h-k-1}}{1-\beta}u}-\frac{1}{1+\frac{1-\beta^{h-k}}{1-\beta}u}\right]}{\left[1-\frac{1}{1+\frac{1-\beta^h}{1-\beta}u}\right]}.
\end{equation*}
As $u$ increases, the numerator is a positive decreasing function, while the denominator is an increasing function of $u$. Hence, the proportion $\frac{\alpha(h,k,u)}{\alpha(h,u)}$ is a decreasing function of $u$ as well.

%
%


\subsubsection{The Wishart Process}

The analysis of risk measured by volatility can be extended to the multivariate framework and volatility-covolatility matrices. This leads to the Wishart Autoregressive (WAR(1)) process that is the multivariate analogue of the ARG(1) process [see Gourieroux, Jasiak, Sufana (2009) for discrete time and Cuchiero et al. (2011) for continuous time]. Due to the matrix framework and the positivity satisfied by a volatility-covolatility matrix, the conditional Laplace transform, is usually written as:
\begin{equation*}
	\Psi(\Gamma,1\vert Y_t) = \mathbb{E}\left[\exp\left(-\Tr\left(\Gamma Y_{t+1}\right)\vert Y_t \right)\right],
\end{equation*}
where $\Gamma$ is a matrix of arguments assumed symmetric positive semi-definite and $\Tr$ denotes the trace operator that sums up the diagonal elements of a square matrix. Then we get:
\begin{corollary}
For the WAR(1) process, the FELD leads to the generic element: 
\begin{equation*}
a(h,k,\Gamma) = \Tr\left\{\left(M^h\right)'\left[\Gamma\left(Id+2\Sigma_{h-k}\Gamma\right)^{-1}-\Gamma\left(Id+2\Sigma_{h-k-1}\Gamma\right)^{-1}\right]M^hY_t\right\}.
\end{equation*}
\end{corollary}
\textbf{Proof:} See Appendix A.2.6. \\

It can be checked that this element reduces to the element in the FELD of the ARG (see Section 4.2.5) in the one dimensional case. 

\section{Illustrations}

We now provide some empirical illustrations of the theory presented in the preceding section. 

\subsection{FEKD for Gaussian VAR(1) Process}

We consider a bivariate VAR(1), $Y_t=(Y_{1,t},Y_{2,t})'$, with autoregressive parameter $\Phi = \begin{bmatrix}
	0.5 & 0.1 \\ 
	0.2 & 0.6 \\		
\end{bmatrix}$. The eigenvalues of matrix $\Phi$ are $\lambda_1=0.7$ and $\lambda_2=0.4$, so the model admits a stationary solution for $(Y_t)$. Let the covariance matrix of $\varepsilon_t$ be given by $\Sigma = \begin{bmatrix}
	1 & 0.2 \\ 
	0.2 & 1 \\  
\end{bmatrix}$, that is, the innovations have unit variance and are positively correlated. \\

\begin{figure}[h]
	\centering
	\includegraphics[width=0.35\linewidth]{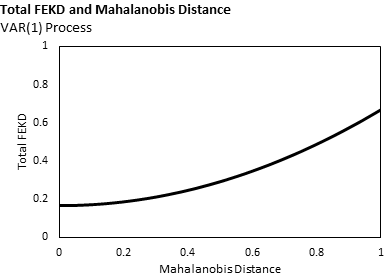}
	\caption{The relationship between the FEKD and the Mahaloanobis distance for the VAR(1) process.}
	\label{fig:fekdgaussian1}
\end{figure} 

The first illustration of interest is how the total FEKD, that is, the left hand side of \eqref{FEKD}, varies with different values of $(Y_t,y)$. In fact, this total relies on how ``far" the value $y$ is from the mean of the conditional predictive density at time $t$, which is equal to $\Phi^hY_t$. To see this, let us define the Mahalanobis distance between $y$ and $\Phi^hY_t$, given by: 
\begin{equation*}
d(y,\Phi^hY_t) = \sqrt{\left(y-\Phi^hY_t\right)'\Sigma_h^{-1}\left(y-\Phi^hY_t\right)},
\end{equation*}
which is an extension of the standard deviation in a multivariate setting. For expository purposes, suppose $Y_t = (2,1)'$ and $h=10$. Then, $\Phi^h(2,1)'=(0.028,0.56)'$. We plot in Figure 1 the relationship between the total FEKD in \eqref{FEKD} and the distance $d(y,\Phi^hY_t)$. The total FEKD increases exponentially with respect to $d(y,\Phi^hY_t)$. This means that values of $y$ further from the mean $\Phi^hY_t$ will have a larger marginal increase in total FEKD. Moreover, note that two points with the same $d(y,\Phi^hY_t)$ will also have the same total FEKD. For example, the points $y=\Phi^h(2,2)'=[0.038, 0.075]'$ and $y=\Phi^h(2,0)'=[0.019, 0.038]'$ both have a Mahalanobis distance of 0.0101 and a total FEKD of 0.1415. However, even if two points share the same Mahalanobis distance, their decompositions can be very different. Let us return to the values $y=\Phi^h(2,2)'=[0.038, 0.075]'$ and $y=\Phi^h(2,0)'=[0.019, 0.038]'$. \\

\begin{figure}[h]
	\centering
	\includegraphics[width=0.65\linewidth]{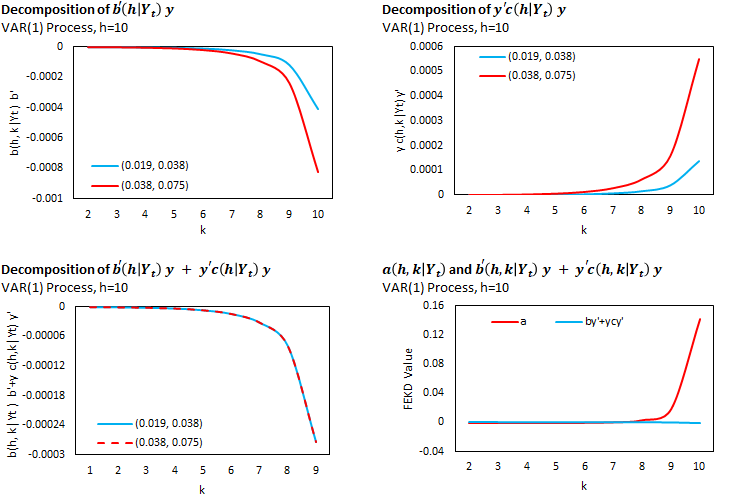}
	\caption{\textit{A comparison of the FEKD for points $y=\Phi^h(2,2)'=[0.038, 0.075]'$ and $y=\Phi^h(2,0)'=[0.019, 0.038]'$, which are near the mean of the predictive density.} }
	\label{fig:fekdgaussian2}
\end{figure}

In the first two graphs of Figure 2, we illustrate the decompositions of $b(h|Y_t)y$ and $y'c(h|Y_t)y$ respectively. It is clear that the point $\Phi^h(2,2)$ has stronger effects on both components. However, when we consider the total sum $b'(h|Y_t)y+y'c(h|Y_t)y$, it is not surprising that both $y=\Phi^h(2,2)'=[0.038, 0.075]'$ and $y=\Phi^h(2,0)'=[0.019, 0.038]'$ produce an identical number, as shown in the bottom left graph. Indeed, since the total FEKD is the same for both points and component $\alpha(h|Y_t)$ is independent of $y$, the sum $b'(h|Y_t)y+y'c(h|Y_t)y$ must also be the same for two points with the same Mahalanobis distance. Furthermore, we note that the comparison of decomposition for these two points is somewhat inconsequential in understanding the dynamics of the VAR(1), since in the bottom right graph, we see that the sum $b'(h,k|Y_t)y+y'c(h,k|Y_t)y$ contributes a very small fraction of the total FEKD in comparison to $a(h,k|Y_t)$. \\

 On the other hand, this latter remark is not always valid for all $y$. Indeed, let us now illustrate an example of points that have a much higher Mahalanobis distance from the mean $\Phi^hY_t$, that is, in the tail ends of the predictive density. For instance, consider  $y=\Phi^h(2,-99)'=[-0.91, -1.83]'$ and $y=\Phi^h(2,101)'=[0.97, 1.94]'$, which have a Mahalanobis distance of 1.0146. \\
 
 \begin{figure}[ht]
 	\centering
 	\includegraphics[width=0.75\linewidth]{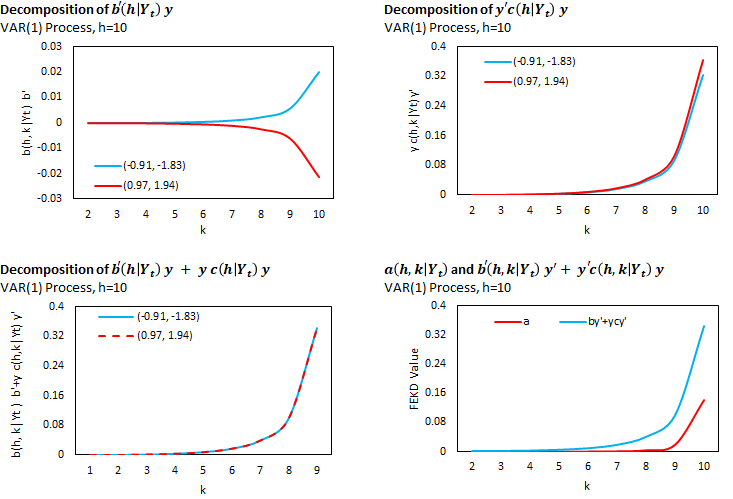}
 	\caption{\textit{A comparison of the FEKD for points $y=\Phi^h(2,-99)'=[-0.91, -1.83]'$ and $y=\Phi^h(2,101)'=[0.97, 1.94]'$, which are at the tail ends of the predictive density.} }
 	\label{fig:fekdgaussian3}
 \end{figure}

In Figure 3, we can see again that the decomposition of components $b'(h|Y_t)y$ and $y' c(h|Y_t)y$ can be very different, even if two points share the same Mahalanobis distance. Also, note that the components of $y' c(h|Y_t)y$ are much larger in magnitude than the components of $b(h|Y_t)y$, which suggests that $y' c(h|Y_t)y$ (the quadratic term) is much more informative on dynamic behaviour in the tails. Furthermore, as seen in the last graph of Figure 3, the sum $b'(h|Y_t)y+y'c(h|Y_t)y$ now takes up a much larger fraction of the total FEKD. Hence, the decomposition is much more sensitive to the choice of $y$ in the tails than it is near the mean.  \\

\subsection{FELD for INAR(1) Model}

Let us consider the INAR(1) process and its FELD derived in Corollary 6, eq. \eqref{FELD_INAR}. The decomposition depends on the current value $Y_t$, on the mean of the Poisson innovation $\lambda$, on the persistence parameter $p$, on the risk aversion $u$, and on the horizon $h$. For expository purposes, we set $Y_t=3$ and $\lambda=2$ in the illustrations, noting that they only enter the FELD as scaling constants and do not change the properties of the decomposition. \\

In Figure 4, we present graphs depicting the relationship between the total FELD and the risk aversion parameter $u$ in $[0.1,2.8]$, under different values of persistence parameter $p$. The lines in each figure represent the FELD for different horizons, with darker shades representing later horizons. For instance, the black line represents the total FELD for horizon 10. \\

\begin{figure}[ht]
	\centering
	\includegraphics[width=0.7\linewidth]{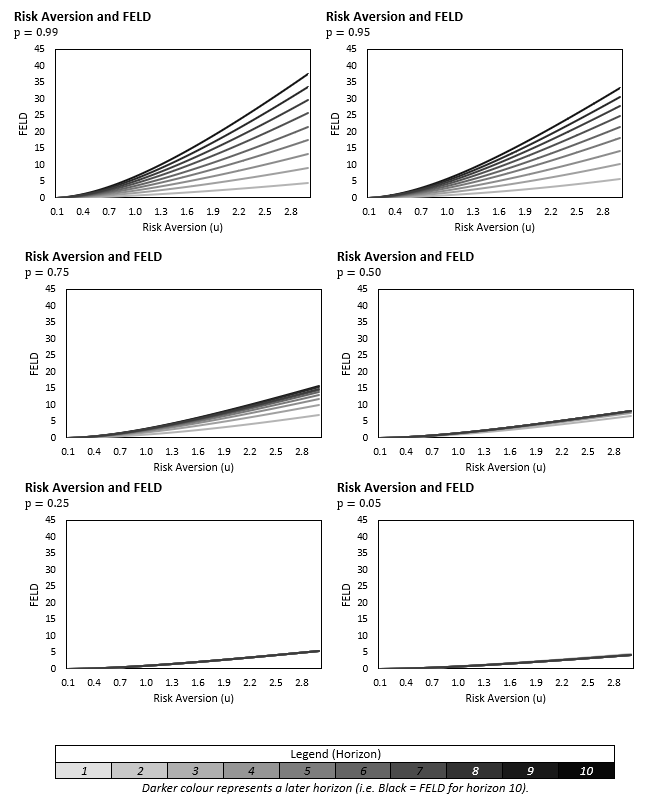}
	\caption{The total FELD for the INAR(1) for each horizon and its relationship with the risk aversion parameter $p$.}
	\label{fig:feldinar1}
\end{figure}

In general, the total FELD is an increasing function of $u$. That is, when a decision maker is more risk averse, the total information gain at a future horizon is also higher, measured by means of the relative change in the Laplace transform of the process. The FELD is also influenced by the level of persistence in the INAR(1). When persistence is low, the information gain in each horizon is similar; in the bottom right graph for $p=0.05$, the line for each horizon is almost stacked on top of each other, whereas the lines in the later horizons are higher in the top left graph for $p=0.99$. Indeed, persistence also influence the steepness of the FELD in each horizon. \\

In Figure 5, we consider decompositions of the FELD at horizons $h=1,...,10$, based on a grid of values in $u=(0.5,1,2)$ and $p=(0.1,0.5,0.95)$. The height of each bar in the graphs correspond to the value of the FELD on the left-hand side of \eqref{FELD_INAR}. They are decomposed into smaller bars with different shades, with the darker shades corresponding to a larger $k$ on the right-hand side of \eqref{FELD_INAR}. \\ 

\begin{figure}[ht]
	\centering
	\includegraphics[width=0.7\linewidth]{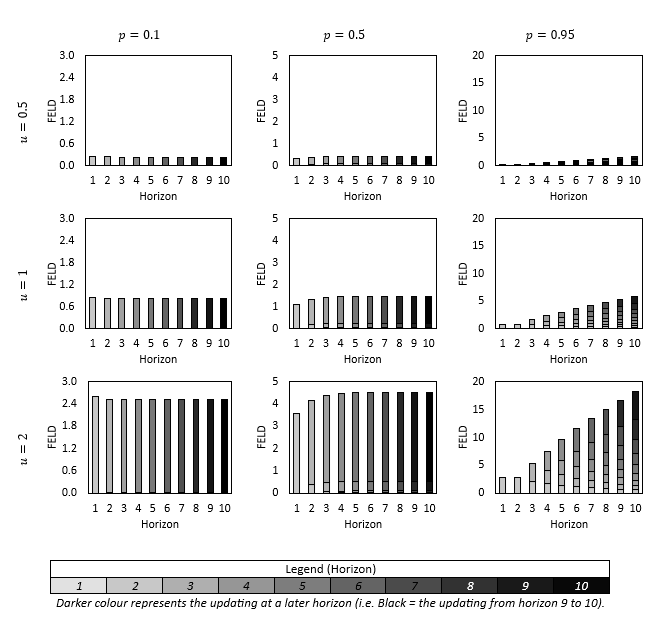}
	\caption{\textit{Total FELD for Horizons 1 – 10 as a function of the Risk Aversion Parameter (u) and persistence parameter (p).}}
	\label{fig:feldinar2}
\end{figure}

The risk aversion parameter only influences the total FELD amount, but it has no impact on the decomposition of its components. Instead, the persistence of the process influnces the FELD in two ways. Firstly, at higher levels of persistence, there is higher contribution of short term updates in previous horizons. For example, let us consider the FELD for horizon 10 at $u=2$ (i.e. the graphs in the third row). When there is almost no persistence ($p=0.1$), the bar at horizon 10 is completely in black, meaning that only the update from horizon 9 to 10 contributes to the total FELD. When persistence is high ($p=0.95$), the bar at horizon 10 contains various shades of grey, which means that updates from previous horizons (e.g. 1 to 2, 2 to 3, 3 to 4, etc.) also contribute to the total FELD. This is also depicted in Figure 6, where the FELD is instead taken as a fraction out of 100. \\ 

\begin{figure}[ht]
	\centering
	\includegraphics[width=0.7\linewidth]{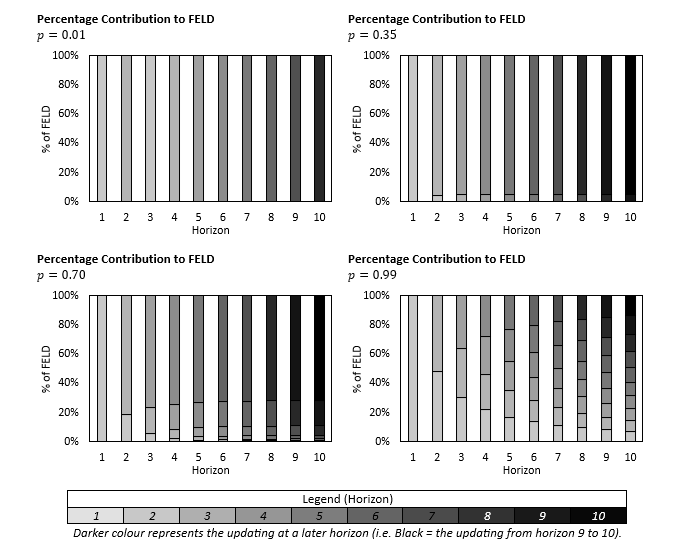}
	\caption{\textit{FELD for INAR(1) processes with varying u and p.}}
	\label{fig:feldinar3}
\end{figure}

 Secondly, for lower levels of persistence, the total FELD seems to ``converge" towards a long-run value. This is due to the stationarity of the INAR(1) process. Indeed, although it is not visible in this illustration, the FELD also converges to a long-run value for $p=0.95$, albeit at a much further horizon. We plot in Table 1 the limiting values of the FELD as $h$ tends to infinity for various combinations of $(p,u)$. To no surprise, the limiting values are increasing in both $u$ and $p$.  \\
 
\begin{table}[ht]
	\centering
	\begin{tabular}{l|llllllllll}
		\hline
			\hline
		 $\rho$ $\backslash$ $u$& 0.1  & 0.4  & 0.7  & 1     & 1.3   & 1.6   & 1.9   & 2.2   & 2.5   & 2.8   \\
			\hline
		0.05 & 0.01 & 0.15 & 0.41 & 0.77  & 1.20  & 1.69  & 2.21  & 2.76  & 3.33  & 3.91  \\
		0.15 & 0.01 & 0.17 & 0.46 & 0.87  & 1.35  & 1.89  & 2.47  & 3.08  & 3.7   & 4.38  \\
		0.25 & 0.01 & 0.19 & 0.52 & 0.98  & 1.53  & 2.14  & 2.80  & 3.50  & 4.22  & 4.97  \\
		0.35 & 0.02 & 0.22 & 0.61 & 1.13  & 1.77  & 2.47  & 3.23  & 4.03  & 4.87  & 5.73  \\
		0.45 & 0.02 & 0.26 & 0.72 & 1.34  & 2.08  & 2.91  & 3.81  & 4.76  & 5.75  & 6.76  \\
		0.55 & 0.02 & 0.31 & 0.87 & 1.63  & 2.54  & 3.56  & 4.66  & 5.82  & 7.03  & 8.27  \\
		0.65 & 0.03 & 0.40 & 1.12 & 2.10  & 3.27  & 4.58  & 6.00  & 7.49  & 9.04  & 10.63 \\
		0.75 & 0.04 & 0.56 & 1.57 & 2.94  & 4.58  & 6.42  & 8.39  & 10.48 & 12.65 & 14.88 \\
		0.85 & 0.06 & 0.93 & 2.62 & 4.90  & 7.63  & 10.69 & 13.99 & 17.47 & 21.09 & 24.81 \\
		0.95 & 0.19 & 2.81 & 7.86 & 14.71 & 22.90 & 32.07 & 41.98 & 52.43 & 63.28 & 74.43\\
		\hline
			\hline
	\end{tabular}		
	\caption{The limiting values of the FELD for combinations of $(\rho,u)$.}
\end{table}

\begin{figure}[h]
	\centering
	\includegraphics[width=0.7\linewidth]{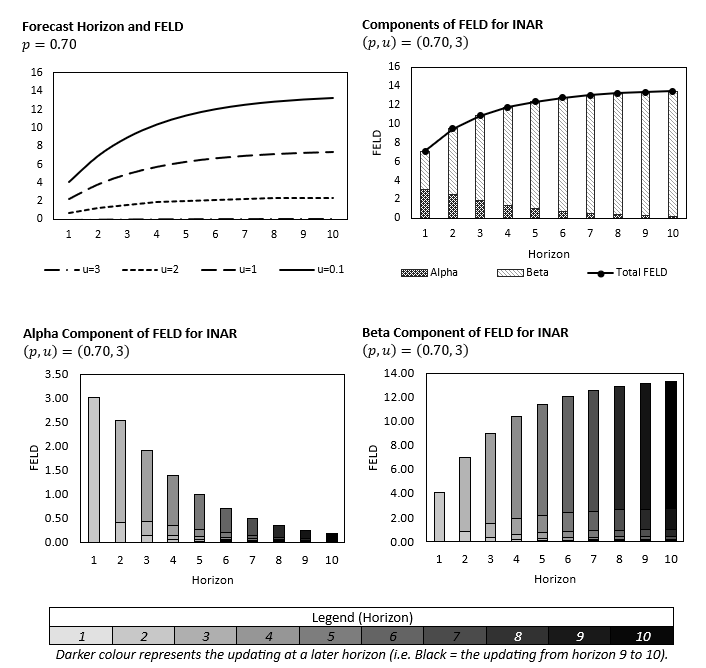}
	\caption{\textit{FELD as a function of the Forecast Horizon, and a decomposition of components in the FELD for parameter combination $(p,u)=(0.70, 3)$.}}
	\label{fig:feldinar4}
\end{figure}

Another decomposition of interest is the separation of the FELD into the time-varying component $\alpha$ and the fixed component $\beta$. This is exemplified in Figure 7 below for the INAR(1) process with $(p,u)=(0.7,3)$. Although the FELD is increasing over horizon $h$, it is the fixed component that plays a gradually larger role in comparison to the time varying component. Indeed, we expect to see this behaviour of the $\alpha$ component, since it has the interpretation of an impulse response function; for stationary processes, the IRF should exhibit transitory behaviour rather than long-run or permanent changes. Looking at a more granular decomposition of the $\alpha$ and $\beta$ componenents themselves, there are no special patterns that differ greatly from the overall FELD. \\

\subsection{FELD for ARG(1) Process}

Let us now illustrate the properties of FELD under the ARG(1) process, with focus on the terms $\alpha(h,u)$ and $\alpha(h,k,u)$. In Figure 8 below, we present the relationship between the total FELD term $\alpha(h,u)$ and risk aversion $u$ for varying levels of parameter $\beta$, which characterizes the serial dependence. \\

\begin{figure}[h]
	\centering
	\includegraphics[width=0.7\linewidth]{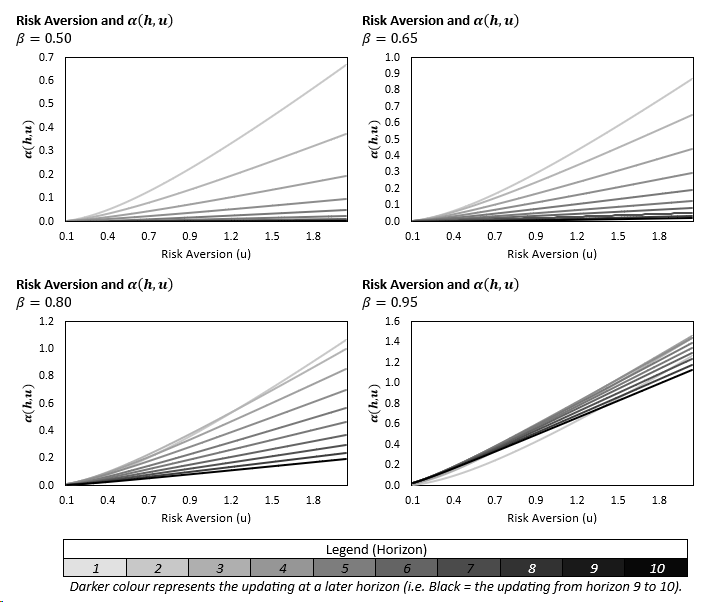}
	\caption{\textit{The relationship between the risk aversion parameter $u$ and the total value of the FELD at different horizons and $\beta$.}}
	\label{fig:feldarg1}
\end{figure}

Let us first consider the cases where $\beta$ is relatively high, say above $\beta = 0.9$. For all horizons $h$, the total FELD term $\alpha(h,u)$ is an increasing function of $u$. However, the rate of increase differs across horizons. In particular, when $u$ is small, the total FELD is higher for the later horizons $h$ (i.e the darker lines are above the lighter ones). This relationship reverses when $u$ is high and there seems to be a point at which the lighter lines ``cross" the darker ones. When $\beta$ is low, this relationship is seemingly absent, and the lighter lines are always higher than the darker ones. However, there is still a ``cross" point for these graphs, but due to the scaling of the axis, they cannot be seen clearly, since they appear at very low values of $u$. Indeed, the value of $\beta$ influences the crossing condition between horizons. In particular, it can be shown that the crossing condition for horizons $h$ and $h+1$ is given by:
\begin{equation*}
 u^*(h,\beta)= \frac{(1-\beta)\left(\beta^{h+1}+\beta^h-1\right)}{\left(1-\beta^h\right)\left(1-\beta^{h+1}\right)}.
\end{equation*}

In the left graph of Figure 9, we depict the curves for $u^*(h,\beta)$ as a function of $\beta$. Each line corresponds to an horizon $h$, with the darker lines representing later horizons. We can see that for smaller $h$, the total FELD will cross with $h+1$ at a much higher level of risk aversion $u$. For instance, at $\beta=0.90$, the curve for $h=1$ will cross with $h=2$ at $u*=3.737$. However, for the curve $h=2$ and $h=3$, the crossing point is at $u^*=1.047$. These values can be seen on the right graph of Figure 9. \\

\begin{figure}[h]
	\centering
	\includegraphics[width=0.7\linewidth]{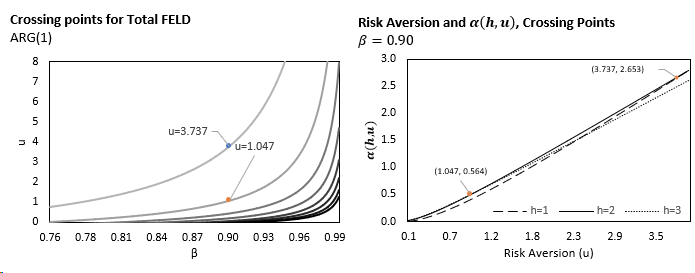}
	\caption{The points $u^*(h,\beta)$ where the total FELD for the ARG(1) process will cross between horizons. }
	\label{fig:feldarg2}
\end{figure}

Finally, we consider the properties of the FELD decomposition for the ARG(1) process. We consider a grid of values on $(u,\beta)$, for $u=0.5,1,3$ and $\beta=0.1,0.5,0.9$ in Figure 10, which shows the fraction $\frac{\alpha(h,k,u)}{\alpha(h,u)}$. There are two main takeaways from this demonstration. Firstly, as the autoregressive parameter $\beta$ increases (i.e going from top to bottom), then the contribution of previous horizons to the current horizon is higher. Secondly, as the risk aversion parameter $u$ increases (i.e. going from left to right), the contribution of previous horizons to the current horizon is lower. Hence, at the bottom left graph, we can see that even for horizon $h=10$, there is significant contribution from $\alpha(h,k,u)$ for values $k=1,...9$. On the other hand, at the top right graph, the bar for horizon $h=10$ is almost completely homogenous in colour. \\

\begin{figure}[h]
	\centering
	\includegraphics[width=1\linewidth]{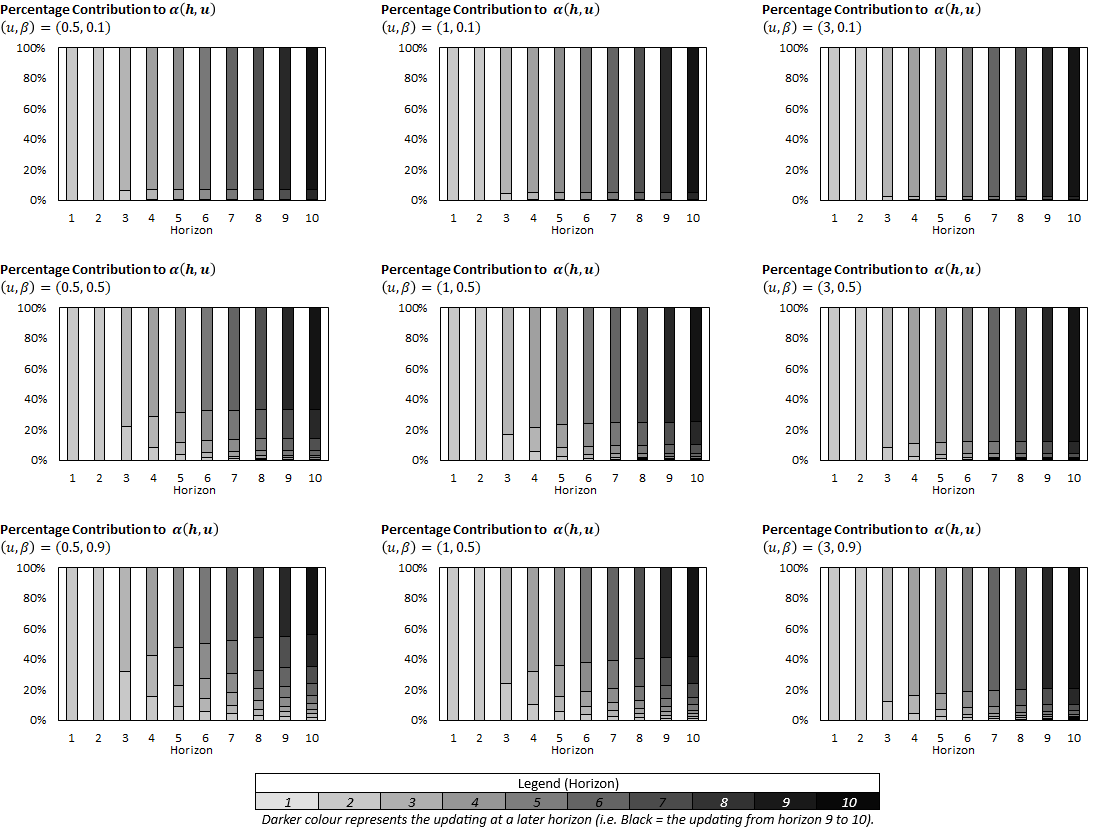}
	\caption{\textit{Percentage contribution of $\alpha(h,k,u)$ to $\alpha(h,u)$ for varying $(u,\beta)$.}}
	\label{fig:feldarg3}
\end{figure}

\section{Statistical Inference}

Let us focus on statistical inference for the FELD, the approach being similar for the FEKD. The decomposition is functional, indexed by the current value $Y_t =y$ and the risk aversion parameter $u$. Then we consider the functional estimator in both the parametric and nonparametric frameworks. 

\subsection{Parametric Dynamic Models}
When the dynamic model is parameterized as in the examples of Section 4.2, the decomposition takes the form:
\begin{equation}
	\gamma(h\vert u,y;\theta) = \sum_{k=1}^{h-1}\gamma (k,h \vert u,y; \theta), 
\end{equation}
where $\theta$ denotes the parameter, $u$ captures the risk aversion and $y$ is the generic value of $Y_t$. If $\hat{\theta}_T$ is a consistent estimator of $\theta$ and is asymptotically normal:
\begin{equation*}
	\sqrt{T}\left(\hat{\theta}_T - \theta\right) \xrightarrow[]{d} N(0,V),
\end{equation*}
the FRED can be estimated by plugging in $\hat{\theta}_T$ in the decomposition: 
\begin{equation*}
	\gamma(h\vert u,y;\hat{\theta}_T) = \sum_{k=1}^{h-1}\gamma (k,h \vert u,y; \hat{\theta}_T), \ \forall \ u,y. 
\end{equation*}
Moreover, uniform confidence bands can be derived by applying the $\delta$-method with respect to $\hat{\theta}_T$. Indeed, the doubly indexed vector  $\text{vec}\  \hat{\gamma}_T(u,y)=\left[\hat{\gamma}_T(k,h,\vert u, y), \ h=1,...,H, \ k=0,...,h-1\right]$ is asymptotically normal. \\

\begin{lemma}
The estimator $\hat{\gamma}_T$ is asymptotically normal such that: 
\begin{equation*}
	\sqrt{T}\left[\hat{\gamma}_T(u,y)-\gamma_0(u,y)\right]\xrightarrow{d} N(0,V''),
\end{equation*}
where $\gamma_0(u,y)$ is computed at the true value $\gamma_0$ and the expansion of the asymptotic variance-covariance matrix is given in Appendix B.1. 
\end{lemma}


\subsection{Nonparametric Dynamic Model}

Nonparametric inference can also be applied under the Markov assumption in the one-dimensional case. Let us for instance consider the FELD. The estimation approach is in four steps:
\begin{enumerate}
	\item Fix a value of the absolute risk aversion $u$ and a maximal horizon $H$. 
	\item Estimate the value of the conditional Laplace transform:
	\begin{equation*}
		\begin{split}
		\Psi(u,h \vert Y_t =y)=&  \mathbb{E}\left[\exp\left(-uY_{t+h}\right)\vert Y_t = y \right], \\
	\text{by its Nadaraya-Watson counterpart:} &	\\
		 \hat{\Psi}_T(u,h \vert Y_t =y)=	& \sum_{t=h+1}^{T}\left[K\left(\frac{y_t-y}{b}\right)\exp\left(-uy_{t+h}\right)\right]\bigg/\sum^{T}_{t=h+1}K\left(\frac{y_t-y}{b}\right),\\ 
		\end{split}
	\end{equation*}
	where $b>0$ is the bandwidth and $K$ a kernel function such that $\int K(u) du = 1$ and $K(-u) = K(u)$ for all values of $u$ in the domain of $K$.
	\item Then the quantity:
	\begin{equation*}
		\log\left[\frac{\Psi(u;h-k\vert Y_{t+k}=y)}{\Psi(u;h-k-1\vert Y_{t+k}=z)}\right]
	\end{equation*}
	can be consistently approximated by:
	\begin{equation*}
		\log\left[\frac{\hat{\Psi}_T(u;h-k\vert y)}{\hat{\Psi}_T(u;h-k-1\vert z)}\right].
	\end{equation*}
	\item A second application of the Nadaraya-Watson approach will provide the estimation of the generic term in the FELD. More precisely, we get:
	\begin{equation*}
		\hat{\gamma}_T(k,h\vert u,y) = \sum_{t=h+1}^{T}\left\{K\left(\frac{y_t-y}{b}\right)	\log\left[\frac{\hat{\Psi}_T(u;h-k\vert y_{t+k})}{\hat{\Psi}_T(u;h-k-1\vert y_{t+k+1})}\right]\right\}\bigg/\sum^{T}_{t=h+1}K\left(\frac{y_t-y}{b}\right).
	\end{equation*}
\end{enumerate}

\section{Application to Cyberrisk}

The prevalance of the internet in our daily lives means that individuals are now able to communicate, transfer and store large amounts of information with just a mobile device. This has significantly improved the way in which businesses operate on a daily basis and has transformed the structure of our modern economy. For example, hospitals have adopted digital patient records which can be accessed by any institution in their network, and alleviated the need to maintain or deliver physical patient files. However, this also offers an opportunity for bad faith actors to intercept or steal information, leading to potentially disastrous outcomes for the victims involved. As such, there is demand for businesses and government agencies to model and quantify the risk of cyber attacks in order to insure against these prospects. In this section, we demonstrate how the FELD can be used in the multivariate Negative Binomial Autoregressive framework in studying the decomposition of frequency in cyber attacks. 

\subsection{The Challenges and the Available Data}

The data on cyber attacks is difficult to obtain. Firstly, there is no consensus on what a cyber attack is. Indeed, there are many definitions, and in a sense it is an umbrella term which includes a variety of illicit behaviours or acts to obtain digital information. Secondly, the collection of data on cyber attacks is rather limited. Although the internet is available almost everywhere today, its accessibility was much less so even just 20 years ago. Hence, tracking cyber attacks has eased only in recent years. Moreover, recording reliable and confirmed cyber attacks is a challenging task, since firms or organizations that are subject to these attacks have an incentive to hide their occurrences. Nonetheless, recent research has appealed to the Privacy Rights Clearinghouse (PRC) dataset, which includes information on publicly reported data breaches across the United States between 2005 and 2022\footnote{https://privacyrights.org/data-breaches. Other cyber databases are Advisen and SAS Oprisk [see Eling, Ibragimov and Ning (2023) for a comparison]}. The PRC was founded in 1992 by the University of San Diego School of Law. The data are gathered from different sources including the Attorney General offices, government agencies, nonprofit websites and media. The reports do not follow a consistent procedure, which may lead to a lack of accuracy and representativeness of the data. Nevertheless, this database is usually employed to analyze cyberrisk [Eling and Loperfido (2017), Eling and Jung (2018), Barati and Yankson (2022), Lu et al. (2024)]. \\

We focus our attention on the modelling of cyber attack frequency counts and adopt the sample used in Lu et al. (2024). The data contains four types of breaches defined as by the PRC:
\begin{itemize}
	\item DISC - Unintended disclosures which do not involve hacking, intentional breaches, or physical losses. 
	\item HACK - Hacked by an outside party or infected by malware. 
	\item INSD - Breach due to an insider, such as an employee, contractor or customer. 
	\item ELET - Lost, discarded or stolen physical devices. 
\end{itemize}
The plots of these four time series are shown in Figure 12 below. 
\begin{figure}[h]
	\centering
	\includegraphics[width=0.75\linewidth]{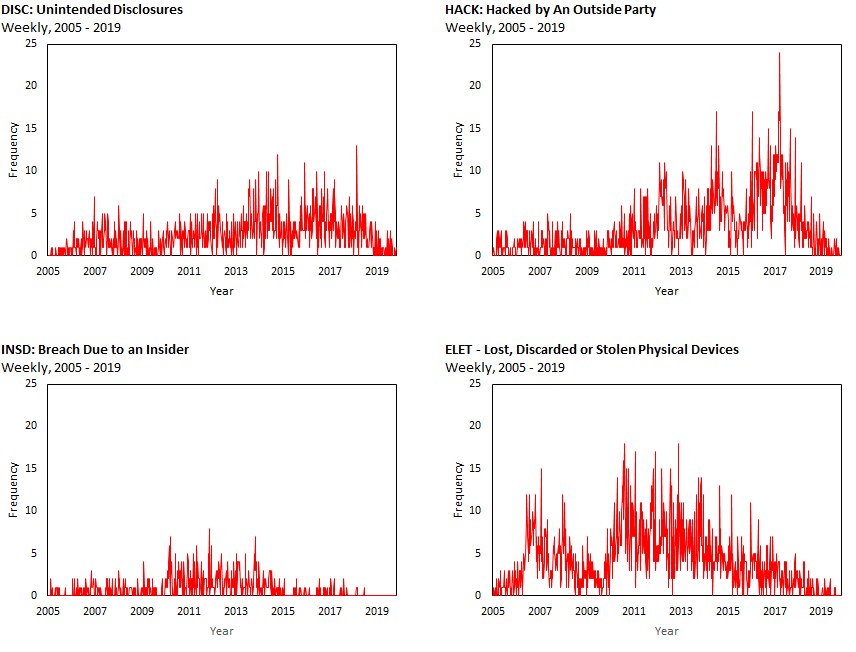}
	\caption{\textit{Time series counts of the four data breaches: DISC, HACK, INSD, ELET.}}
	\label{fig:cyberdata}
\end{figure}
All four series feature a large number of zero counts, with 146, 138, 440 and 108 instances for DISC, HACK, INSD and ELET respectively. They also have varying levels of occurrences; for instance, while HACK and INSD can both be considered breaches with malicious intent, the latter occurs much less frequently on average. This is due to the fact that it is much easier for an outsider to gain access without getting caught (such as using an IP spoofer), than an insider to attempt a breach. To gain insights on the distribution of each series we provide summary statistics and density plots of each type of breach below. \\
\begin{table}[h]
	\centering
	\begin{tabular}{l|llll}
		\hline
		\hline
		Process & Mean & Variance & Excess Skewness & Excess Kurtosis \\ \hline
		\hline
		DISC    & 2.41 & 4.55     & 1.26            & 2.17            \\
		HACK    & 3.28 & 11.13    & 1.60            & 3.31            \\
		INSD    & 0.78 & 1.46     & 2.10            & 5.42            \\
		ELET    & 4.08 & 12.28    & 1.17            & 1.41            \\ \hline
		\hline
	\end{tabular}
	\caption{Summary statistics of data set.}
\end{table}

In each series, the variance is much larger than the mean, which suggests that the data exhibit overdispersion. There is also positive excess skewness, which means that each count is skewed towards the right; this is not surprising since there are a large concentration of zeroes for each type of breach. Furthermore, all the series exhibit excess kurtosis, so they have relatively fatter tails compared to the normal distribution. \\

\begin{figure}[h]
	\centering
	\includegraphics[width=0.6\linewidth]{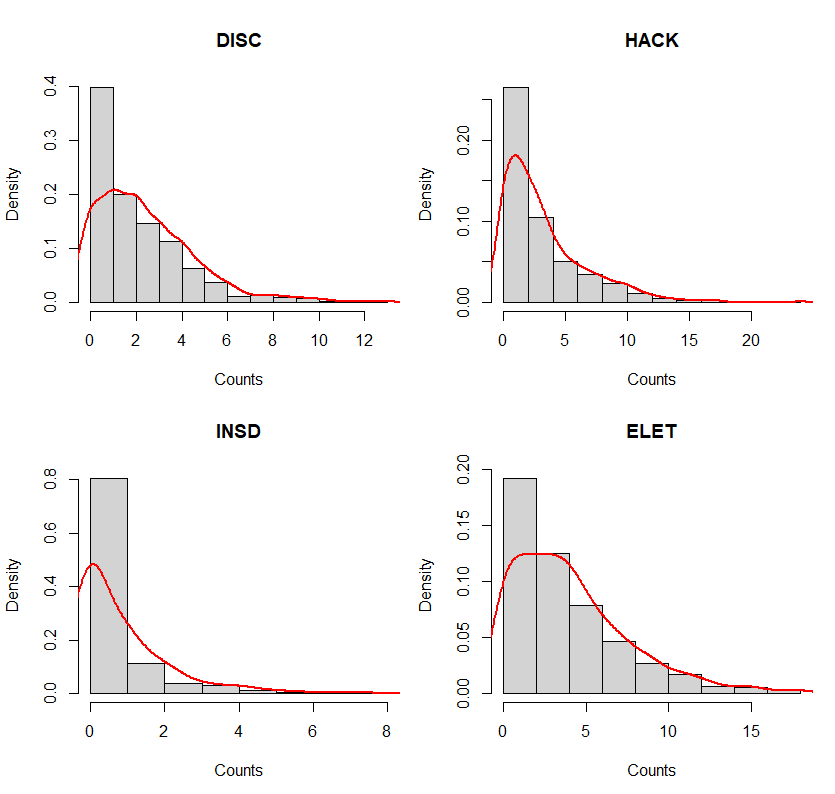}
	\caption{\textit{Histogram and corresponding density plot of the four series.}}
	\label{fig:density}
\end{figure}


%
%

%
%
%

\subsection{Univariate Analysis}

The INAR model of Section 4.2.4. is the basic dynamic model for a series of count data. However, it implies a marginal Poisson distribution, for which the mean is equal to the variance. Therefore, it is not compatible with data featuring overdispersion, that is, where the variance is much larger than the mean, as seen in Table 2. Consequently, the risk can be underestimated in such a framework. The INAR model can be extended for more flexibility by introducing stochastic intensity. This leads to the univariate Binomial Autoregressive Process (NBAR) [see Gourieroux, Lu (2019)]. We present the univariate model in this section which will be used in this analysis (the bivariate NBAR will be presented in Section 7.3.1.).

\subsubsection{Univariate NBAR}

The process is defined by its state space representation with a state variable $X_t$ interpretable as stochastic intensity. This representation is as follows:
\begin{enumerate}
	\item \textbf{Measurement Equation - } Conditional on $\underline{X}_{t+1}$,$\underline{Y}_{t}$, the count process $Y_{t+1}$ is assumed to be $\mathcal{P}(\beta X_{t+1})$ with $\beta>0$. 
	\item \textbf{Transition Equation -} Conditional on $\underline{X}_t$,$\underline{Y}_t$, the stochastic intensity factor $X_{t+1}$ is assumed to be a centred gamma distribution $\gamma(\delta+Y_t,\beta,c)$, with shape parameter $\delta+Y_t$ and scale parameter $c$.
\end{enumerate} 
In total, there are three parameters to be estimated: $\beta$, $\delta$ and $c$. More specifically, $\beta$ characterizes not only the serial dependence of the process, but also the level of conditional overdispersion. The process described above can be represented by the causal chain:
\begin{equation*}
	... X_t \rightarrow Y_{t+1} \rightarrow X_{t+1} \rightarrow Y_{t+2} ...
\end{equation*}
that is a network with one hidden layer and one hidden neuron. Note that the NBAR process is an affine process with conditional Laplace transform [Gourieroux, Lu (2019), Proposition 2]:
\begin{equation}
	\begin{split}\label{lptunbar}
	\Psi(u,h|I_t) & = \mathbb{E}\left[\exp(-uY_{t+h})\vert Y_{t}\right]\\
	& = \frac{\left[1+\beta c_{h-1}(1-\exp(-u))\right]^{Y_t}}{\left[1+\beta c_{h}(1-\exp(-u))\right]^{\delta+Y_t}},\\
	& = \exp\left[-a^{(h)}(u)Y_t+b^{(h)}(u)\right],\\
	\end{split}
\end{equation}
where:
\begin{equation*}
	\begin{split}
		a^{\circ h }(u) & =-\log\left[1+\beta c_{h-1}(1-\exp(-u))\right]+\log\left[1+\beta c_{h}(1-\exp(-u))\right],\\
		b^{\circ h }(u) & =- \delta \log\left[1+\beta c_{h}(1-\exp(-u))\right],\\
	\end{split}
\end{equation*}
and $\rho = \beta c$ and the sequence $(c_h)$ defined by $c_h=c \frac{1-\rho^h}{1-\rho}$ or equivalently, $\beta c_h = \rho \frac{1-\rho^h}{1-\rho}$. This process is strictly stationary if $\rho<1$ and its stationary distribution is obtained for $h$ tending to infinity. The associated Laplace transform is: 
\begin{equation}\label{unclptunnbar}
	\begin{split}
		c(u)=\Psi(u,\infty|I_t) & = \frac{1}{\left[1+\beta c(1-\exp(-u))\right]^\delta},\\
	\end{split}
\end{equation}
and corresponds to the negative binomial distribution $NB(\delta,\rho)$.

\subsubsection{Estimation}

The NBAR process is a nonnegative Markov process where the distribution is characterized by its conditional Laplace transform at horizon 1. This transition depends on the parameters $\beta$, $c$, $\delta$, by means of $\rho = \beta c$ and $\delta$ only. Therefore, the parameters $\rho = \beta c$ and $\delta$ are identifiable [Gourieroux and Lu (2019)], but not $\beta$ and $c$ separately. Then, we can consider the following two estimation methods: \\

\textbf{(1) Linear Regression} \\ 

By the law of iterative expectation, it can be shown that: 
\begin{equation*}
	\mathbb{E}(Y_{t}|Y_{t-1})=\mathbb{E}\left[\mathbb{E}(Y_t|X_{t-1})|Y_ {t-1}\right]=\rho Y_{t-1} + \rho\delta.
\end{equation*}
Therefore, the parameters $\rho$ and $\rho\delta$ (resp. $\delta$) can be estimated using OLS via a regression of $Y_{t}$ on $Y_{t-1}$. However, this method is not asymptotically efficient due to the presence of conditional heteroscedasticity in the NBAR process.\\

\textbf{(2) Maximum Likelihood Estimation} \\

The likelihood function of the NBAR process is given by [Gourieroux, Lu (2019), eq. 4.1]: 
\begin{equation}
	\log \ell(\theta) = \sum_{t=2}^T \log p(Y_t|Y_{t-1};\rho,\delta),
\end{equation}
where:
\begin{equation}
	p(Y_t|Y_{t-1};\rho,\delta) = \frac{\rho^{Y_t}\Gamma(\delta+Y_t+Y_{t-1})}{Y_t!\Gamma(\delta+Y_{t-1})(1+\rho)^{\delta+Y_{t}+Y_{t-1}}},
\end{equation}
and $\Gamma$ denotes the gamma function. Unlike OLS, the MLE estimator is asymptotically efficient. We present below the estimates for $\delta$ and $\rho$ and their associated standard errors under the OLS and MLE methods. 


\begin{table}[h]
	\centering
	\begin{tabular}{c|cc|cc|cc|cc}
		\hline
		\hline 
	\textbf{Parameter}& \textbf{DISC}& &  \textbf{HACK} && \textbf{INSD}& & \textbf{ELET}& \\
		 & OLS & MLE &  OLS & MLE & OLS & MLE & OLS & MLE \\
		\hline
		$\delta$  &4.9514	&3.2244	&1.8481& 1.6917	&1.3857&0.9745	& 3.9177&2.0913	\\
					& (1.0208)	& (0.3990)	& (0.3167)&(0.1683)	& (0.3697)& (0.1314)& (0.6399)	&(0.2273)	\\
		\hline
		$\rho$	  & 0.3278	& 0.4282	& 0.6400 & 0.6601	& 0.3619 & 0.4464	& 0.5107 & 0.6616\\
		&(0.0341)	& (0.0325)	& (0.0277)&(0.0336)	& (0.0280)& (0.0398)	&(0.0310)& (0.0287)	\\
		\hline
		\hline
	\end{tabular}
	\caption{OLS and MLE estimates of $\rho$ and $\delta$.}
\end{table} 
As expected, the estimates of the parameter $\rho$ are positive, to capture the overdispersion, and smaller than one, that is compatible with the stationarity of the count processes. Moreover, the largest persistences are for HACK and ELET. We also observe that the OLS estimates of the persistence (overdispersion) parameter $\rho$ are always smaller than their ML counterparts. This provides some insight on the finite sample bias due to the omission of the conditional heteroscedasticity in the OLS approach. 

\subsubsection{Univariate FELD}
\begin{corollary}
	For the univariate NBAR process described in Section 7.2.1, the FELD is given by: 
	\begin{equation}\label{FELD_NBAR_UNI}
\begin{split}
	& \left\{u\rho^h+\log\left[1+\beta c_{h-1}(1-\exp(-u))\right]-\log\left[1+\beta c_{h}(1-\exp(-u))\right]\right\}Y_t\\ 
	& +\delta \rho u - \delta \rho^{h+1} u - \delta \log\left[1+\beta c_{h}(1-\exp(-u))\right] \\
	= & \sum_{k=0}^{h-1}\left\{\rho^{k+1}\log\left[\frac{1+\beta c_{h-k-1}(1-\exp(-u))}{1+\beta c_{h-k-2}(1-\exp(-u))}\right]-\rho^k\log\left[\frac{1+\beta c_{h-k}(1-\exp(-u))}{1+\beta c_{h-k-1}(1-\exp(-u))}\right]\right\}Y_t\\
	& -\delta \rho\left\{\log\left[\frac{1+\beta c_{h-k}(1-\exp(-u))}{1+\beta c_{h-k-1}(1-\exp(-u))}\right]-\log\left[\frac{1+\beta c_{h-k-1}(1-\exp(-u))}{1+\beta c_{h-k-2}(1-\exp(-u))}\right]\right\} \\
	& -\delta \rho \left\{\rho^{k}\log\left[\frac{1+\beta c_{h-k}(1-\exp(-u))}{1+\beta c_{h-k-1}(1-\exp(-u))}\right]-\rho^{k+1}\log\left[\frac{1+\beta c_{h-k-1}(1-\exp(-u))}{1+\beta c_{h-k-2}(1-\exp(-u))}\right]\right\}\\
	& - \delta \log\left[\frac{1+\beta c_{h-k}(1-\exp(-u))}{1+\beta c_{h-k-1}(1-\exp(-u))}\right], \ \ \forall \ u > 0,\\
\end{split}
	\end{equation}
	and $\beta c_h = \rho \frac{1-\rho^h}{1-\rho}$. 
\end{corollary}

\textbf{Proof:} See Appendix B.2.1. \\

We now present the results of the FELD using the MLE estimated NBAR models for each of the four series for varying risk aversion $u$ and horizon $h$. We focus on the term:
\begin{equation*}
	\left\{u\rho^h+\log\left[1+\beta c_{h-1}(1-\exp(-u))\right]-\log\left[1+\beta c_{h}(1-\exp(-u))\right]\right\},
\end{equation*}
which captures the marginal effect of one additional breach $Y_t$ on the Total FELD. 
\begin{figure}[h]
	\centering
	\includegraphics[width=0.75\linewidth]{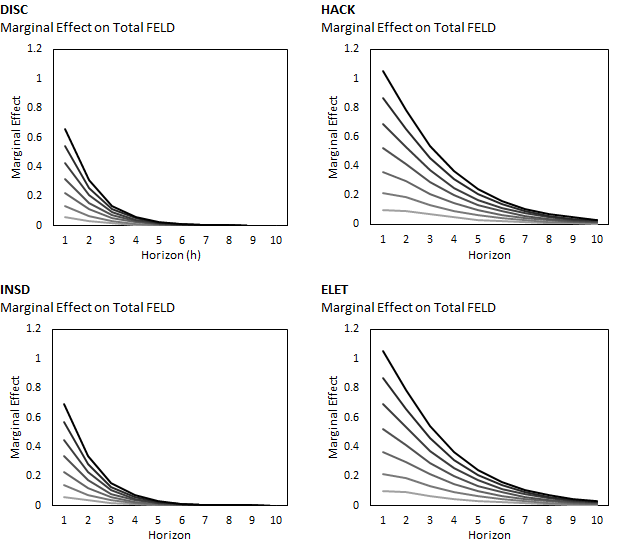}
	\caption{\textit{The marginal effect of an additional breach on the Total FELD.}}
	\label{fig:feldnbar}
\end{figure}
In each graph, the marginal effect of $Y_t$ is plotted against the forecast horizon $h$. As $h$ falls, the marginal effect of the history $Y_t$ diminishes exponentially. The darker lines represent higher risk aversion parameter $u$. Thus, higher risk aversion means a higher marginal effect of $Y_t$. Intuitively, this implies that if the firm is more risk averse, an additional breach that occurs today will yield higher uncertainty in the forecast (measured by the marginal effect on the Total FELD). Moreover, the marginal effects seem to be higher for HACK and ELET, and lower for DISC and INSD. As seen in Section 3.4, the decompositions in Figure 13 also have interpretation in terms of spot and forward values of derivatives written on the number of cyber events. This is related to the literature on the pricing of cyberinsurance contracts [Fahrenwaldt et al. (2018)].

\subsection{Bivariate Analysis}

The four series correspond to different types of cyber operational risks. Among them two of the risks correspond to breaches with malicious intent, i.e. HACK and INSD. We will focus on the joint analysis of these two risks\footnote{It is possible to study the four risks together, but with more complex models [see online appendix B3 for the extension to the $K$-dimensional NBAR.]}, which has to account for both cross-sectional and serial dependencies between the series. A first insight on the cross-sectional dependence is through the joint stationary distributions. This is illustrated in Figure 14, where on the left hand side we provide a plot of values $(Y_{1,t},Y_{2,t})$ and on the right hand side a plot of the associated Gaussian ranks ($\pi(Y_{1,t})$,$\pi(Y_{2,t})$), i.e. the estimated copula after the Gaussian transform\footnote{The raw data are first ranked by increasing order. Rank$(Y_{j,t})$, the rank of $Y_{j,t}$, is valued in $[1,...,T]$. Then, $\pi(Y_{j,t})=\Phi^{-1}\left[\text{Rank}(Y_{j,t})/T\right]$, where $\Phi$ is the cumulative distribution function of the standard normal distribution.}\footnote{This Gaussian transformed unconditional copula is completed on the count of occurrences. It differs from a copula completed from losses that account for severities [see Eling, Jung (2018)].}.  \\

\begin{figure}[h]
	\centering
	\includegraphics[width=1\linewidth]{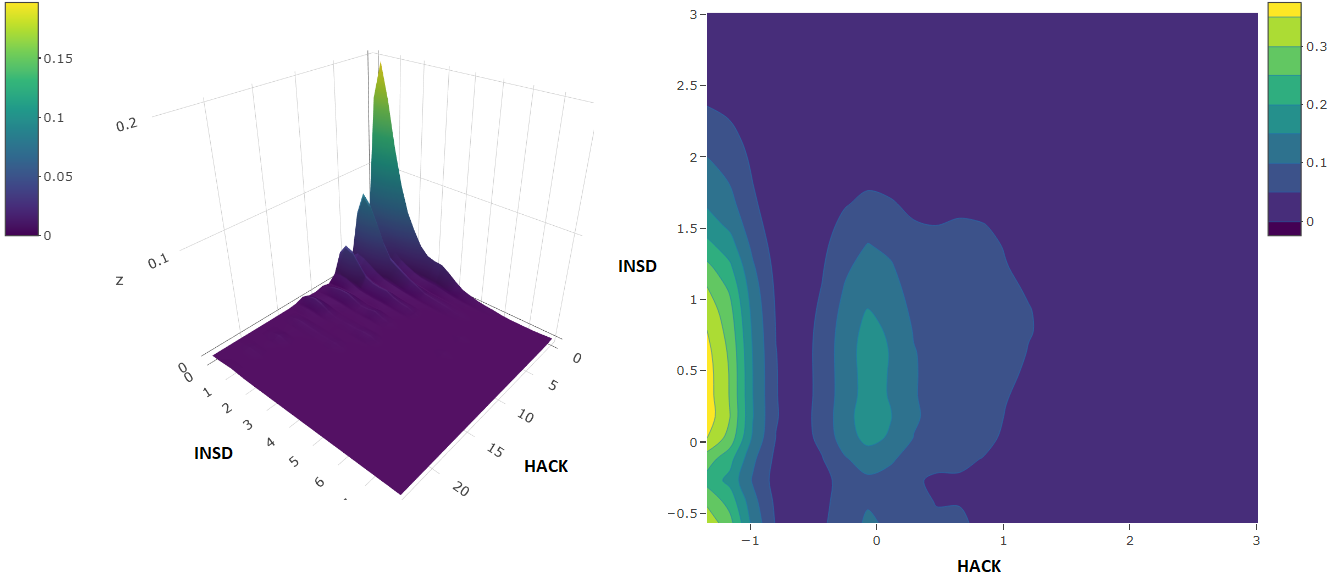}
	\caption{\textit{The joint density plot of HACK and INSD (left) and the contour density plot of their associated Gaussian ranks (right).}}
	\label{fig:gaussianranks}
\end{figure}

Such plots are usually given for continuous variables or for count variables taking a large number of values. In our framework, the discrete nature of the variable has to be taken into account in the interpretation. For instance, we observe an overweighting of zero counts for the HACK variable, which can be seen in the left plot of Figure 14, where the joint density rests ``against" its axis. The right panel in Figure 14 also shows that the two series features some right tail independence.

\subsubsection{Bivariate NBAR}

Let us denote $Y_{1t},Y_{2t}$ the two count series. We can now introduce three state variables interpretable as specific and common stochastic intensities, denoted $X_{1,t},X_{2,t}$ and $Z_t$, respectively. The nonlinear state space representation becomes:
\begin{enumerate}
	\item  \textbf{Measurement Equation -} Conditional on $\underline{Y}_t$, $\underline{X}_{t+1}$, $\underline{Z}_{t+1}$, the variables $Y_{1,t+1}$, $Y_{2,t+1}$ are independent $Y_{j,t+1} \sim \mathcal{P}(\alpha_jZ_{t+1}+\beta_jX_{j,t+1})$ for $j=1,2$. 
	\item \textbf{Transition Equations -} Conditional on $\underline{Y}_{t}$, $\underline{X}_{t}$, $\underline{Z}_{t}$, the variables $X_{1,t+1}$, $X_{2,t+1}$, $Z_{t+1}$ are independent such that $X_{j,t+1} \sim \gamma(\delta_j+Y_{j,t})$, for $j=1,2,$ and $Z_{t+1} \sim \gamma(\delta+\sigma_1Y_{1,t}+\sigma_2Y_{2,t},0,c)$. 
\end{enumerate}
The dimensionality and the presence of stochastic intensity factors now lead to 9 parameters to be estimated: $\alpha_1$, $\alpha_2$, $\beta_1$, $\beta_2$, $\delta_1$, $\delta_2$, $\sigma_1$, $\sigma_2$, and $\delta$. The process described above corresponds to a more complicated causal scheme with one hidden layer and three hidden neurons.
\begin{equation*}
	\begin{tikzcd}
		Y_{1,t}\arrow[r] \arrow[rd] & X_{1,t+1}\arrow[r] & Y_{1,t+1}\\
		& Z_{t+1} \arrow[ru] \arrow[rd]&\\
		Y_{2,t}\arrow[r]\arrow[ru] &  X_{2,t+1}\arrow[r] & Y_{2,t+1}
	\end{tikzcd}
\end{equation*}

It is easily checked that the bivariate NBAR model is an affine model. Its conditional Laplace transform at horizon $1$ is given by [see Appendix A.2.7]:
\begin{equation}
\begin{split}\label{lptbnbar}
\Psi(u,1|Y_t)=\mathbb{E}\left[\exp(-u'Y_{t+1})\vert Y_{t}\right]&= \exp\left[-a_1(u_1,u_2)Y_{1,t}-a_2(u_1,u_2)Y_{1,t}-b(u_1,u_2)\right],\\
\end{split}
\end{equation}
where: 
\begin{equation*}
	\begin{split}
		a_1(u_1,u_2)= &  \log\left[1+\beta_1(1-\exp(-u_1))\right]+\sigma_1\log[1+\alpha_1(1-\exp(-u_1))+\alpha_2(1-\exp(-u_2))],\\
		a_2(u_1,u_2)= &  \log\left[1+\beta_2(1-\exp(-u_2))\right]+\sigma_2\log[1+\alpha_1(1-\exp(-u_1))+\alpha_2(1-\exp(-u_2))],\\
		b(u_1,u_2) =&  \delta_1\log[1+\beta_1(1-\exp(-u_1))]+\delta_2\log[1+\beta_2(1-\exp(-u_2))]\\
		& + \delta\log[1+\alpha_1(1-\exp(-u_1))+\alpha_2(1-\exp(-u_2))].\\ 
	\end{split}
\end{equation*}
We see from the expression of the conditional Laplace transform that all parameters are identifiable. 

\subsubsection{Estimation}

The increase of the parameter dimension and the introduction of a common stochastic intensity lead to a larger number of identifiable parameters to be estimated, equal to 9. A first estimation to consider is by applying a Vector Autoregressive (VAR) representation based on the linear prediction formula [see Appendix A.2.8]: 
\begin{equation}\label{VAR_NBAR}
\mathbb{E}\left[Y_{t}|Y_{t-1}\right] = \begin{bmatrix}
	\alpha_1\delta + \beta_1\delta\\
	\alpha_2\delta + \beta_2\delta\\ 
\end{bmatrix}+\begin{bmatrix}
	\alpha_1\sigma_1+\beta_1  & \alpha_1\sigma_2 \\
	\alpha_2\sigma_1 &  \alpha_2\sigma_2 +\beta_2 \\
\end{bmatrix}\begin{bmatrix}
	Y_{1,t-1}\\
	Y_{2,t-1}
\end{bmatrix}.
\end{equation}
Using OLS to estimate the model $Y_t = C + AY_{t-1} +\varepsilon_t$, we obtain the following estimates: \\
\begin{table}[h]
	\centering
	\begin{tabular}{ccc}
		\hline
		\hline
		&    $Y_{1,t}$   &   $Y_{2,t}$   \\
		\hline
	$Y_{1,t-1}$	&  0.638     &   0.005   \\
	&  (0.028)    &   (0.012)  \\
	$Y_{2,t-1}$	&  0.053      &   0.361   \\
	&  (0.076)      &   (0.034)   \\
	$Constant$	&  1.145     &  0.486  \\ 
	&  (0.140)     & (0.062)     \\
	\hline
	\hline
	\end{tabular}
\end{table}

The corresponding eigenvalues of $\hat{A} = \begin{bmatrix}
	0.638 & 0.053 \\
	0.005 & 0.361 \\
\end{bmatrix}$  are $\lambda_1 = 0.639$ and $0.360$, which suggests that the process is stationary in the conditional mean.\\

 More generally, we can apply a Method of Moments (MM) procedure based on the unconditional pairwise moment restrictions of the form: 
\begin{equation}\label{mm_nbar}
	\begin{split}
	\mathbb{E}\left\{\left[\exp(-u'Y_t)-\Psi(u,1|Y_t)\right]\exp(-v'Y_{t-1})\right\} = 0,
	\end{split}
\end{equation}
valued for different $u$ and $v$, by considering the orthogonality between prediction errors and past values. A global estimation of the 9 parameters can be based on the unconditional pairwise moments given in \eqref{mm_nbar} after selecting at least 9 quadruples $(u_1,u_2,v_1,v_2)$ of linearly independent moment restrictions\footnote{We choose quadruplets which correspond to a range of different risk aversion scenarios. For instance, the quadruplet (0.41,0.01,0.41,0.01) corresponds to the scenario where there is high risk aversion on only on the series $Y_{1,t}$. Likewise, the quadruplet (0.41,0.41,0.01,0.01) reflects the case where there is high risk aversion for both series, but only at time $t$.}. We also include 6 additional moment conditions implied by the OLS first order conditions. Applying a generalized method of moment procedure for the 15 moment conditions, we obtain the following estimates displayed in Table 4.  \\

\begin{table}[h]
\centering
	\begin{tabular}{ccccccccc}
		\hline
				\hline
	$\alpha_1$	& $\alpha_2$ & $\beta_1$ & $\beta_2$    & $\delta_1$  &  $\delta_2$& $\sigma_1$ & $\sigma_2$ & $\delta$   \\
			\hline
0.118	& -0.067 & 0.647 & 0.391 & 1.20 &  1.27 & -0.075  & 0.453 & 1.492 \\
(0.277)& (0.367) & (0.078) & (0.226) & (0.626) & (0.998) & (0.374) & (1.155) & (0.671) \\ 
				\hline
			\hline
	\end{tabular}
	\caption{Estimated values of the 9 coefficients in the bivariate model of HACK and INSD.}
\end{table}

These estimates imply $	\hat{\hat{C}} = \begin{bmatrix}
	1.143\\
		0.485 \\
	\end{bmatrix} \ \hat{\hat{A}} = \begin{bmatrix}
	0.701 & 0.053 \\
		0.005 & 0.361 \\
	\end{bmatrix}$. Hence, our choice of quadruplets not only captures different risk aversion scenarios, but also produce somewhat agreeable estimates with that of OLS.  We also note that the estimated values suggest that the process is stationary\footnote{The stationary conditions are given by: $1-\alpha_1-\sigma_1\beta_1>0$, $1-\alpha_2-\sigma_2\beta_2>0$ and $(1-\alpha_1-\sigma_1\beta_1)(1-\alpha_1-\sigma_1\beta_1)>\sigma_1\sigma_2\beta_1\beta_2$ [Gourieroux, Lu (2019), Proposition 3].}.

\subsubsection{FELD for Bivariate NBAR}

A closed form expression for the conditional Laplace transform at horizon $h$ is difficult to obtain in closed form. However, the dynamic affine property of the bivariate NBAR means that it can be derived numerically by means of recursion. In particular, we have:
\begin{equation}\label{recursive_nbar}
	\begin{split}
		\Psi(u,h|Y_t)=\mathbb{E}\left[\exp(-u'Y_{t+h})\vert Y_{t}\right]&= \exp\left[-a_1^{(h)}(u_1,u_2)Y_{1,t}-a_2^{(h)}(u_1,u_2)Y_{1,t}-b^{(h)}(u_1,u_2)\right],\\
	\end{split}
\end{equation}
where: 
\begin{equation*}
	\begin{split}
		a_1^{(h)}(u_1,u_2)= & a_1(a_1^{(h-1)}(u_1,u_2),a_2^{(h-1)}(u_1,u_2)),\\
		a_2^{(h)}(u_1,u_2)= & a_2(a_1^{(h-1)}(u_1,u_2),a_2^{(h-1)}(u_1,u_2)),\\
				b^{(h)}(u_1,u_2)= & b(a_1^{(h-1)}(u_1,u_2),a_2^{(h-1)}(u_1,u_2)), \ \forall\ h \geq 2.
	\end{split}
\end{equation*}
The estimated FELD for the model can still be obtained by applying the recursion in \eqref{recursive_nbar} to generate the terms $\Psi(u,h-k|I_{t+k})$ in \eqref{FELD} and plugging in the estimated values for the 9 parameters from Table 4. We consider two exercises with varying values of $(u_1,u_2)$ [i.e. the risk aversion parameter for HACK and INSD] and of $(Y_{1,t},Y_{2,t})$ [i.e. the last observed historical value for HACK and INSD] to showcase the FELD for the data. 

\begin{table}[h]
	\begin{tabular}{lll}
		\hline
		\hline
		\multicolumn{3}{l}{\textbf{Exercise 1:} Risk Aversion and Historical Counts are the SAME for both HACK and INSD } \\
		\hline
		\hline
	Low Risk Aversion, Low Historical Count	&  $(u_1,u_2)=(0.5, 0.5)$     &  $(Y_{1,t},Y_{2,t})=(0,0)$    \\
	High Risk Aversion, Low Historical Count	&  $(u_1,u_2)=(2,2)$     &  $(Y_{1,t},Y_{2,t})=(0,0)$    \\
	Low Risk Aversion, High Historical Count	&  $(u_1,u_2)=(0.5,0.5)$     &  $(Y_{1,t},Y_{2,t})=(5,5)$    \\
	High Risk Aversion, High Historical Count	&  $(u_1,u_2)=(2,2)$     &  $(Y_{1,t},Y_{2,t})=(5,5)$    \\
	\hline
	\hline
	\end{tabular}
\end{table}

The first exercise is to see how the risk aversion and the historical counts of the two cyberrisks influence the FELD. We compare four scenarios: [1] Low Risk Aversion and Low Historical Counts. [2] High Risk Aversion and Low Historical Counts. [3] Low Risk Aversion and High Historical Counts. [4] High Risk Aversion and High Historical Counts. The results are presented in Figure 15 below. \\

\begin{figure}[h]
	\centering
	\includegraphics[width=1\linewidth]{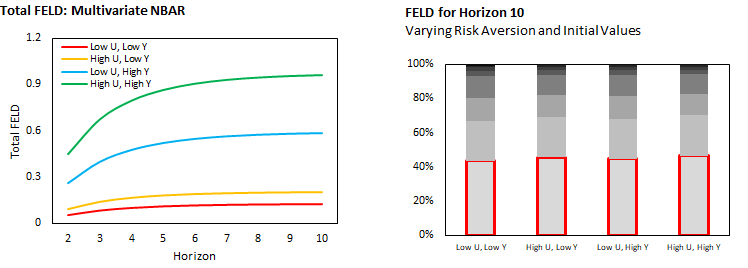}
	\caption{\textit{A comparison of four scenarios where risk aversion and historical counts are the same for both HACK and INSD.}}
	\label{fig:F15}
\end{figure}

The graph on the left hand side depicts the total FELD up to horizon 10. We observe two important properties of the model. First, the total FELD is increasing in $u$, since the red and blue lines are lower than the green and yellow ones. Second, the total FELD is increasing in $Y$, since the green and blue lines are higher than the red and yellow ones. Indeed, recall that the FELD is a measure of risk. If a firm were to observe high historical counts $Y$ of cyberattacks, then the risk of future attacks should be higher. Likewise, when observing the same level of historical counts of cyberattacks, there is a higher implied risk for a more risk averse firm. \\

On the right hand side graph, we show the decomposition of the total FELD at horizon 10. For ease of comparison between the four scenarios, we express the FELD as a percentage of the total. We have highlighted in red the contribution of risk in updating between horizon 1 and horizon 2 on the FELD at horizon 10. Although there are four different scenarios, we can see that the percentage contribution is roughly 40\% in all the cases. This suggests that if we are equally risk averse on the two types of cyber risk, and observe similar levels of historical counts, then our decomposition of the total risk is insensitive to the various levels. \\

Of course, the assumption that the two series have the same risk aversion parameters or the same observed historical values is unrealistic. Hence, we consider a second exercise where these values for HACK and INSD. We compare four new scenarios: [1] Low/High Historical Count for HACK/INSD, High/Low Risk Aversion for HACK/INSD. [2] High/Low Historical Count for HACK/INSD, High/Low Risk Aversion for HACK/INSD.  [3] Low/High Historical Count for HACK/INSD, Low/High Risk Aversion for HACK/INSD.  [4] High/Low Historical Count for HACK/INSD, Low/High Risk Aversion for HACK/INSD. The results are presented in Figure 16.\\

\begin{table}[h]
	\begin{tabular}{lll}
		\hline
		\hline
		\multicolumn{3}{l}{\textbf{Exercise 2:} Different Values of Risk Aversion and Historical Counts for HACK and INSD} \\
		\hline
		\hline
		Low $Y_{1,t}$, High $u_1$,	High $Y_{2,t}$, Low $u_2$	&  $(u_1,u_2)=(2,0.5)$     &  $(Y_{1,t},Y_{2,t})=(0,5)$    \\
		High $Y_{1,t}$, High $u_1$,	Low $Y_{2,t}$, Low $u_2$		&  $(u_1,u_2)=(2,0.5)$     &  $(Y_{1,t},Y_{2,t})=(5,0)$    \\
		Low $Y_{1,t}$, Low $u_1$,	High $Y_{2,t}$, High $u_2$		&  $(u_1,u_2)=(0.5,2)$     &  $(Y_{1,t},Y_{2,t})=(0,5)$    \\
		High $Y_{1,t}$, Low $u_1$,	Low $Y_{2,t}$, High $u_2$		&  $(u_1,u_2)=(0.5,2)$     &  $(Y_{1,t},Y_{2,t})=(5,0)$    \\
		\hline 
		\hline
	\end{tabular}
\end{table}

\begin{figure}[h]
	\centering
	\includegraphics[width=1\linewidth]{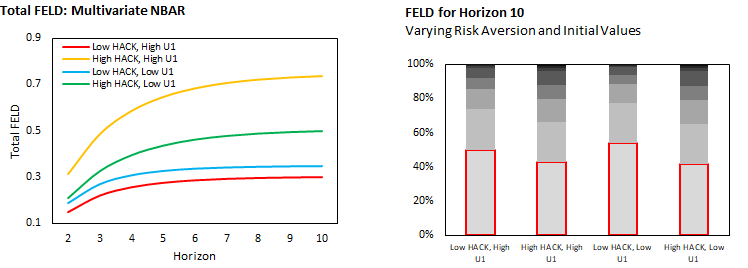}
	\caption{\textit{A comparison of four scenarios where risk aversion and historical counts are different for both HACK and INSD.}}
	\label{fig:F16}
\end{figure}

On the left hand side graph of Figure 16, the yellow and green lines correspond to a high observed historical value for HACK, but a low historical value observed for INSD. These lines are much higher than the blue and red lines, which correspond to the opposite case. Intuitively, this means that observing a high count of breaches due to hacking from an outside party will generate more uncertainty than observing a high count of insider breaches. This is a compelling result since for a firm, it is easier to diagnose the security leak and intitiate preventive measures for future insider breaches than it is for hacking from an outside party. A more straightforward observation is that for high levels of HACK, a lower risk aversion parameter means less uncertainty [i.e. the yellow line is higher than the green one]. \\

On the right hand side graph of Figure 16, we show the decomposition of the total FELD at horizon 10 for this exercise. A notable difference is that now, the decomposition is quite different for the four scenarios. In particular, when updating from horizon 1 to 2 (outlined in red), the scenarios have different contributions to the FELD at horizon 10. When the observed historical count for HACK is low and INSD is high (that is, the first and third bars in the graph), the risk is front loaded since the risk of updating between horizon 1 and 2 accounts for over 40\% of the FELD in these cases. On the other hand, when a high historical count of HACK and a low count of INSD is observed (that is, the second and fourth bars in the graph), the risk is has more spread across future horizons. This means that a high count of observed insider breaches implies a more front loaded risk.\\

The exercises above demonstrate two important factors. Firstly, by considering two series together in a bivariate model, we are able to take advantage of the cross-sectional dependencies between the two types of cyber attacks, even if our decomposition measure is a dynamic separation of effects.  Secondly, the FELD we have proposed in this paper has as many decompositions as there are $u$ and $Y_t$. In particular, we see that in exercise 1 and exercise 2, the decompositions can be very different and dependent on risk aversion and observed historical counts. For an insurance firm that specializes in cyber risk, these types of scenario analysis can help price their insurance products, or allocate resources efficiently to hedge against future cyber attacks for a range of customers.

\section{Concluding Remarks}

In this paper, we have introduced decomposition formulas for the analysis of global forecast errors in nonlinear dynamic models. These formulas are based on functional measures of nonlinear forecast with respect to either their transition densities in the Forecast Error Kullback Decomposition, or the conditional log-Laplace transform in the Forecast Error Laplace Decomposition. These measures and their decompositions are extensions of the Forecast Error Variance Decomposition (FEVD) used in the linear dynamic framework. This latter decomposition is for global shocks on the current value of the process of interest, without trying to define (identify) the sources of the global shocks. The advantage is that this measure and its decomposition with respect to horizon and updating are identifiable, as are also the FEKD and FELD. Such decompositions could be extended to functional measures of economic or financial interest such as conditional Lorenz curves used in inequality analysis or conditional quantiles [Montes-Rojas (2019)]. This is left for future research. 

\newpage

\section{References}

Al-Osh, M. and A., Azaid (1987): ``First-Order Integer Valued Autoregressive (INAR(1)) Processes", Journal of Time Series Analysis, 8, 261-275. \\

Arrow, K. (1965): ``Aspects of the Theory of Risk Bearing", The Theory of Risk Aversion, Helsinki, Reprinted in ``Essays in the Theory of Risk Bearing", Markham Publishing Co., Chicago, 1971, 90-109.\\ 

Barati, M. and B., Yankson (2022): ``Predicting the Occurrence of Data Breach," International Journal of Information Management Data Insights, 2. \\

Cebula, J., and L. Young (2010): ``A Taxonomy of Operational Cyber Security Risks", Technical Note, Software Engineering Institute, Carnegie Mellon University.\\

Chauvet, M., and S. Potter (2005): ``Forecasting Recession Using the Yield Curve", Journal of Forecasting, 24, 77-103. \\ 

Cox, J., Ingersoll, J., and S., Ross (1985): ``A Theory of the Term Structure of Interest Rates", Econometrica, 53, 385-407. \\

Cuchiero, C., Filipovic, D., Mayerhofer, E. and J., Teichmann (2011): ``Affine Processes on Positive Semi Definite Matrices", The Annals of Applied Probability, 21, 397-463. \\ 

Darolles, S., Gourieroux, C., and J., Jasiak. (2006): ``Structural Laplace Transform and Compound Autoregressive Models". Journal of Time Series Analysis, 27, 477-503. \\

Doob, J. (1953): ``Stochastic Processes", Wiley and Sons. \\

Duffie, D., Filipovic, D., and W., Schachermayer. (2003): ``Affine Processes and Applications in Finance". The Annals of Applied Probability, 13, 984-1053. \\ 

Eling, M., Ibragimov, R., and D., Ning, (2023): ``Time Dynamics of Cyber Risk", University of St. Gallen. \\

Eling, M., and K., Jung (2018): ``Copula Approaches for Modelling Cross-Sectional Dependence of Data Breach Losses", Insurance: Mathematics and Economics, 82, 167-180. \\

Eling, M., and N. Loperfido (2017): ``Data Breaches: Goodness of Fit Pricing and Risk Management", Insurance: Mathematics and Economics, 75, 126 - 136. \\

Embrechts, P. (2000): ``Actuarial versus Financial Pricing of Insurance", J. Risk Finance, 1, 17-26. \\

Estrella, A., and F. Mishkin (1998): ``Predicting US Recessions: Financial Variables as Leading Indicators", Review of Economics and Statistics, 80, 45-61. \\ 

Fahrenwaldt, M., Weber, S., and K., Weske (2018): ``Pricing of Cyber Insurance Contracts in a Network Model", ASTIN Bulletin, 48, 1175-1218. \\

Feller, W. (1971). ``An Introduction to Probability Theory and Its Applications", (2nd Edition), Vol 2, Wiley, New York. \\ 

Fishburn, R., and R., Vickson (1978): ``Theoretical Foundations of Stochastic Dominance", in Stochastic Dominance: An Approach to Decision Making Under Risk, Whitmore, G. and Findlay, M. (eds). DC Health, London.\\

Gourieroux, C., and J., Jasiak (2006): ``Autoregressive Gamma Processes", Journal of Forecasting, 25, 129-152. \\

Gourieroux, C., Jasiak, J., and R., Sufana (2009): ``The Wishart Autoregressive Process of Multivariate Stochastic Volatility", Journal of Econometrics, 150, 167-181. \\ 

Isakin, M., and P., Ngo (2020): ``Variance Decomposition Analysis for Nonlinear Economic Models". Oxford Bulletin of Economics and Statistics, 82, 1362-1374.\\ 

Kauppi, H., and P., Saikkonen (2008): ``Predicting US Recessions with Dynamic Binary Response Models", Review of Economics and Statistics, 90, 777-796. \\

Lanne, M., and H., Nyberg. (2016): ``Generalized Forecast Error Variance Decomposition for Linear and Nonlinear Multivariate Models". Oxford Bulletin of Economics and Statistics, 78(4), 595-603.\\

Lu, Y., Zhang, J., and W., Zhu (2024): ``Cyber Risk Modelling: A Discrete Multivariate Count Process Approach," forthcoming, Scandinavian Actuarial Journal.\\ 

Markowitz, H. (1952): ``Portfolio Selection", The Journal of Finance, 7,77-91. \\

Markowitz, H. (2000): ``Mean-Variance Analysis in Porfolio Choice and Capital Markets", Wiley and Sons, Vol. 66. \\

McKenzie, E. (1985): ``Some Simple Models for Discrete Variate Time Series", Water Resources Bulletin, 21, 645-650. \\

Montes-Rojas, G. (2019): ``Multivariate Quantile Impulse Response Functions", Journal of Time Series Analysis, 40, 730-752. \\

Muirhead, R. (1982): ``Aspects of Multivariate Statistical Theory", Wiley and Sons. \\

Pratt, J. (1964): ``Risk Aversion in the Small and in the Large", Econometrica, 32, 129-136. \\

Rothschild, M., and J., Stiglitz (1970): ``Increasing Risk: I. A Definition", Journal of Economic Theory, 2, 225-243. \\ 

Schwartz, G. and S. Sastry (2014): ``Cyber-insurance framework for large scale interdependent networks", in Proceedings of the 3rd International Conference on High Confidence Networked Systems, 145-154, New-York, The Association on Computing Machinery. \\ 

Sun, H., Xu. M., and P., Zhao (2021): ``Modelling Malicious Hacking Data Breach Risks", North American Actuarial Journal, 25, 484-502. \\

Vickson, R. (1975): ``Stochastic Dominance for Decreasing Absolute Risk Aversion", Journal of Financial and Quantitative Analysis, 10, 799-811. \\

Whittle, P. (1963): ``Prediction and Regulation", Bowman, new edition in 1983. \\

\newpage

\appendix

\section{Proofs}

\subsection{Examples for FEKD}

\subsubsection{Proof of Proposition 3: Gaussian VAR(1)}

The conditional distributions of $Y_{t+h}$ given $Y_{t+k+1}$ and $Y_{t+h}$ given $Y_{t+k}$ are $N(\Phi^{h-k-1}Y_{t+k+1},\Sigma_{h-k-1})$ and $N(\Phi^{h-k}Y_{t+k},\Sigma_{h-k})$ respectively. Hence:
\begin{equation*}
	\begin{split}
		\log\left[\frac{f(y,h-k|I_{t+k})}{f(y,h-k-1|I_{t+k+1}}\right] = & \frac{1}{2}\log\left[\frac{\det\Sigma_{h-k-1}}{\det\Sigma_{h-k}}\right]+\frac{1}{2}\left[y-\Phi^{h-k-1}Y_{t+k+1}\right]'\Sigma_{h-k-1}^{-1}\left[y-\Phi^{h-k-1}Y_{t+k+1}\right] \\ 
		& -\frac{1}{2}\left[y-\Phi^{h-k}Y_{t+k}\right]'\Sigma_{h-k}^{-1}\left[y-\Phi^{h-k}Y_{t+k}\right] \\
		\end{split}
\end{equation*}
Let $A= \left[y-\Phi^{h-k-1}Y_{t+k+1}\right]$. Then, by conditioning on $I_t$, we get: 
\begin{equation*}
	\begin{split}
	&	\frac{1}{2}\mathbb{E}\left\{\left[y-\Phi^{h-k-1}Y_{t+k+1}\right]'\Sigma_{h-k-1}^{-1}\left[y-\Phi^{h-k-1}Y_{t+k+1}\right]\bigg \vert Y_{t}\right\} \\ 
	=	&  \frac{1}{2}\mathbb{E}\left[A'\Sigma_{h-k-1}^{-1}A\vert Y_{t}\right] \\ 
	 = &  \frac{1}{2} \textup{Tr}\left[\Sigma^{-1}_{h-k-1}\mathbb{E}(AA'\vert Y_{t})\right]\\ 
	 = &  \frac{1}{2} \textup{Tr}\left[\Sigma^{-1}_{h-k-1}\left(\mathbb{E}\left(A|Y_t\right)\mathbb{E}\left(A|Y_t\right)'+\mathbb{V}(A\vert Y_{t})\right)\right]\\ 
	 = & \frac{1}{2} \Tr\left\{\Sigma_{h-k-1}^{-1}\left[(y-\Phi^{h}Y_t)(y-\Phi^{h}Y_t)'+\Phi^{h-k-1}\Sigma_{k+1}(\Phi^{h-k-1})'\right]\right\}. \\
	\end{split}
\end{equation*}
Similarly, if $B= \left[y-\Phi^{h-k}Y_{t+k}\right]$, then:
\begin{equation*}
	\begin{split}
		&\frac{1}{2}\mathbb{E}\left[B'\Sigma_{h-k}^{-1}B\vert Y_{t}\right] \\
		=&\frac{1}{2} \Tr\left\{\Sigma_{h-k}^{-1}\left[(y-\Phi^{h}Y_t)(y-\Phi^{h}Y_t)'+\Phi^{h-k}\Sigma_{k}(\Phi^{h-k})'\right]\right\}. \\  
	\end{split}
\end{equation*}

Then, the expression for $\gamma(k,h|I_t)$ is:
\begin{equation*}
	\begin{split}
	\gamma(k,h|I_t) = & \frac{1}{2}\log\left[\frac{\det\Sigma_{h-k-1}}{\det\Sigma_{h-k}}\right]-\frac{1}{2}\Tr\left[\left(\Sigma_{h-k}^{-1}-\Sigma_{h-k-1}^{-1}\right)(y-\Phi^{h}Y_t)(y-\Phi^{h}Y_t)'\right]  \\ 
	& + \frac{1}{2}\Tr\left[\Sigma_{h-k-1}^{-1}\Phi^{h-k-1}\Sigma_{k+1}(\Phi^{h-k-1})'-\Sigma_{h-k}^{-1}\Phi^{h-k}\Sigma_{k}(\Phi^{h-k})'\right]\\
	= & \frac{1}{2}\log\left[\frac{\det\Sigma_{h-k-1}}{\det\Sigma_{h-k}}\right] + \frac{1}{2}\Tr \left[\Sigma^{-1}_{h-k-1}\Phi^{h-k-1}\Sigma_{k+1}\left(\Phi^{h-k-1}\right)'-\Sigma^{-1}_{h-k}\Phi^{h-k}\Sigma_k\left(\Phi^{h-k}\right)'\right] \\ 
	& -\frac{1}{2}Y_t'\left(\Phi^h\right)'\left(\Sigma^{-1}_{h-k}-\Sigma^{-1}_{h-k-1}\right)\Phi^hY_t\\
&	+ Y_t'\left(\Phi^h\right)'\left(\Sigma^{-1}_{h-k}-\Sigma^{-1}_{h-k-1}\right)y - \frac{1}{2}y'\left(\Sigma^{-1}_{h-k}-\Sigma^{-1}_{h-k-1}\right)y, \\
	\end{split}
\end{equation*}
which is quadratic in $y$. 

\subsubsection{Proof of Proposition 4: Binary Markov Chain}
We have: 
	\begin{equation*}
	\begin{split}
		\gamma(k,h|Y_t)= & \mathbb{E}\left\{\log\left[\pi + \lambda^{h-k}(Y_{t+k}-\pi)\right]-\log\left[\pi + \lambda^{h-k-1}(Y_{t+k+1}-\pi)\right]\bigg \vert Y_t\right\}\\
		= & \mathbb{E}\left\{\log \left[\pi + \lambda^{h-k}(1-\pi)\right]Y_{t+k}+\log\left[\pi\left(1-\lambda^{h-k}\right)\right](1-Y_{t+k}) \right. \\
		 & \left.-\log\left[\pi + \lambda^{h-k-1}(1-\pi)\right]Y_{t+k+1}-\log\left[\pi\left(1-\lambda^{h-k-1}\right)\right](1-Y_{t+k+1})\bigg \vert Y_t \right\} \\
		 =& \log\left[\frac{1-\lambda^{h-k}}{1-\lambda^{h-k-1}}\right] + \log\left[\frac{\pi+\lambda^{h-k}(1-\pi)}{\pi(1-\lambda^{h-k})}\right]\left[\pi+\lambda^k(Y_t-\pi)\right] \\
		& -  \log\left[\frac{\pi+\lambda^{h-k-1}(1-\pi)}{\pi(1-\lambda^{h-k-1})}\right]\left[\pi+\lambda^{k+1}(Y_t-\pi)\right]. \\
	\end{split}
\end{equation*}

\subsubsection{Proof of  Corollary 1 : FEVD for Markov Chain }

\begin{equation*}
	\begin{split}
		& \mathbb{V}\left\{\mathbb{E}\left[Y_{t+h}|I_{t+k+1}\right]-\mathbb{E}\left[Y_{t+h}|I_{t+k}\right]\bigg \vert I_t \right\} \\ 
		= & \mathbb{V}\left\{\lambda^{h-k-1}(Y_{t+k+1}-\pi)-\lambda^{h-k}(Y_{t+k}-\pi)\bigg \vert Y_t \right\}\\ 
		= & \mathbb{V}\left\{\lambda^{h-k-1}Y_{t+k+1}-\lambda^{h-k}Y_{t+k}\bigg \vert Y_t \right\}\\
		= & \lambda^{2(h-k-1)}\mathbb{V}\left[Y_{t+k+1}\vert Y_t\right] + \lambda^{2 (h-k)}\mathbb{V}\left[Y_{t+k}\vert Y_t\right] - 2\lambda^{2(h-k)+1}Cov(Y_{t+k+1},Y_{t+k}|Y_t)\\
	\end{split}
\end{equation*}
Note that:
\begin{equation*}
	\begin{split}
	\mathbb{V}\left[Y_{t+k}\vert Y_t\right]=&  \mathbb{E}(Y_{t+k}|Y_t)-\left[\mathbb{E}(Y_{t+k}|Y_t)\right]^2 \\
	=& \mathbb{E}(Y_{t+k}|Y_t)\left[1-\mathbb{E}(Y_{t+k}|Y_t)\right] \\
	= & \left[\pi + \lambda^{k}(Y_t-\pi)\right]\left[1-\pi - \lambda^{k}(Y_t-\pi)\right],\\
	\mathbb{V}\left[Y_{t+k+1}\vert Y_t\right]=& \left[\pi + \lambda^{k+1}(Y_t-\pi)\right]\left[1-\pi - \lambda^{k+1}(Y_t-\pi)\right],\\ 
	\end{split}
\end{equation*}
and
\begin{equation*}
	\begin{split}
	Cov(Y_{t+k+1},Y_{t+k}\vert Y_t) & = \mathbb{E}\left(Y_{t+k+1}Y_{t+k}\vert Y_t\right)-\mathbb{E}\left(Y_{t+k+1}\vert Y_t\right)\mathbb{E}\left(Y_{t+k}\vert Y_t\right)\\ 
	& = \mathbb{E}\left[\mathbb{E}\left(Y_{t+k+1}Y_{t+k}\vert Y_{t+k}\right)\vert Y_t\right]-\mathbb{E}\left(Y_{t+k+1}\vert Y_t\right)\mathbb{E}\left(Y_{t+k}\vert Y_t\right)\\
	& = \mathbb{E}\left[\mathbb{E}\left(Y_{t+k+1}|Y_{t+k}\right) Y_{t+k}|Y_t\right]-\mathbb{E}\left(Y_{t+k+1}\vert Y_t\right)\mathbb{E}\left(Y_{t+k}\vert Y_t\right)\\
	& = \mathbb{E}\left[\left(\pi + \lambda Y_{t+k}-\lambda\pi\right) Y_{t+k}|Y_t\right]-\mathbb{E}\left(Y_{t+k+1}\vert Y_t\right)\mathbb{E}\left(Y_{t+k}\vert Y_t\right)\\
	& = \mathbb{E}\left[\pi Y_{t+k} + \lambda Y^2_{t+k}-\lambda\pi Y_{t+k} \vert Y_t \right]-\left[\pi+\lambda^{k+1}(Y_t-\pi)\right]\left[\pi+\lambda^{k}(Y_t-\pi)\right]\\
		& = \left[\pi+\lambda(1-\pi)\right]\left[\pi+\lambda^{k}(Y_t-\pi)\right]-\left[\pi+\lambda^{k+1}(Y_t-\pi)\right]\left[\pi+\lambda^{k}(Y_t-\pi)\right]\\
		& = \left[\pi+\lambda^{k}(Y_t-\pi)\right]\left[\lambda(1-\pi) - \lambda^{k+1}(Y_t-\pi)\right]\\
		& = \lambda\left[\pi+\lambda^{k}(Y_t-\pi)\right]\left[1-\pi - \lambda^{k}(Y_t-\pi)\right] \\
	\end{split}
\end{equation*}
Thus we have:
\begin{equation*}
\begin{split}
&	\lambda^{2(h-k-1)}\mathbb{V}\left[Y_{t+k+1}\vert Y_t\right] + \lambda^{2 (h-k)}\mathbb{V}\left[Y_{t+k}\vert Y_t\right] - 2\lambda^{2(h-k)-1}Cov(Y_{t+k+1},Y_{t+k}|Y_t) \\
	= & 
	\lambda^{2(h-k-1)} \left[\pi + \lambda^{k+1}(Y_t-\pi)\right]\left[1-\pi - \lambda^{k+1}(Y_t-\pi)\right]+ \lambda^{2(h-k)} \left[\pi + \lambda^{k}(Y_t-\pi)\right]\left[1-\pi - \lambda^{k}(Y_t-\pi)\right]\\ 
	& - 2\lambda^{2(h-k)}\left[\pi+\lambda^{k}(Y_t-\pi)\right]\left[1-\pi - \lambda^{k}(Y_t-\pi)\right] \\ 
		= & 
	\lambda^{2(h-k-1)} \left[\pi + \lambda^{k+1}(Y_t-\pi)\right]\left[1-\pi - \lambda^{k+1}(Y_t-\pi)\right]- \lambda^{2(h-k)} \left[\pi + \lambda^{k}(Y_t-\pi)\right]\left[1-\pi - \lambda^{k}(Y_t-\pi)\right]\\
		 	 = & \pi(1-\pi)\left[\lambda^{2(h-k-1)}-\lambda^{2(h-k)}\right] - \pi(Y_t-\pi)\left[\lambda^{2(h-k-1)}\lambda^{k+1}-\lambda^{2(h-k)}\lambda^{k}\right]\\  
		 	 & + (1-\pi)(Y_t-\pi)\left[\lambda^{2(h-k-1)}\lambda^{k+1}-\lambda^{2(h-k)}\lambda^{k}\right] - (Y_t-\pi)^2\left[\lambda^{2(h-k-1)}\lambda^{2(k+1)}-\lambda^{2(h-k)}\lambda^{2k}\right]\\
		 	 = & \pi(1-\pi)\lambda^{2(h-k-1)}\left[1-\lambda^2\right]- \pi(Y_t-\pi)\lambda^{2h-k-1}\left[1-\lambda\right]+ (1-\pi)(Y_t-\pi)\lambda^{2h-k-1}\left[1-\lambda\right]\\ 
		 	 		 	 = & \pi(1-\pi)\lambda^{2(h-k-1)}\left[1-\lambda^2\right]- (Y_t-\pi)\lambda^{2h-k-1}\left[1-2\pi\right]\left[1-\lambda\right]\\ 
\end{split}
\end{equation*}

\subsubsection{Proof of Proposition 5: Markov Chain}

The $h$-step forward prediction of the Markov Chain, $\mathbb{E}\left[Y_{t+h}|Y_{t}\right]$, is given by:
\begin{equation*}
\begin{split}
		& \mathbb{E}\left[Y_{t+h}|Y_{t}\right] = \mathbb{E}\left[X_{t+h}|X_{t}\right] \\
	& = P^{h}X_t,\\ 
\end{split}
\end{equation*}
and its $h$-step transition density, $f(y,h|I_t)$, is given by:
\begin{equation*}
\begin{split}
		f(y,h|I_t) &= P(Y_{t+h}=y|Y_t) \\ 
	&=\sum_{i=1}^k P(Y_{t+h}=y|Y_t=i)X_{it} \\ 
	&=\sum_{i=1}^k p^{(h)}_{iy}X_{it} \\
	&= P_y^{h}X_{t},\\ 
\end{split}
\end{equation*}
where $P_y^{h}$ denotes the $y$-th row of $P^{h}$. For the FEKD we are interested in the quantity: 
\begin{equation*}
	\begin{split}
	\log f(y,h|I_t) & = \log P(Y_{t+h}=y|Y_{t}) \\ 
	& =\sum_{i=1}^n \log P(Y_{t+h}=y|Y_t=i)X_{it}\\ 
		& =\sum_{i=1}^n \log\left[p^{(h)}_{yi}\right]X_{it}.\\ 
	\end{split}
\end{equation*}
Let us denote $\widetilde{\log}(A)$ as a matrix whose elements are the logged elements of matrix $A$. It follows that: 
\begin{equation*}
	\log P(Y_{t+h}=y|Y_{t}) = \left[\widetilde{\log}(P^{h})\right]_yX_t.
\end{equation*}
Hence, the right-hand side term in the FEKD is given by:
\begin{equation}
	\begin{split}
	\gamma(h,k|I_{t}) & = \mathbb{E}\left[\log P(Y_{t+h}=y|Y_{t+k})-\log P(Y_{t+h}=y|Y_{t+k+1})\vert _t\right]\\
& = \mathbb{E}\left[\left[\widetilde{\log}(P^{h-k})\right]_yX_{t+k}\vert X_t\right]-\mathbb{E}\left[\left[\widetilde{\log}(P^{h-k-1})\right]_yX_{t+k+1}\vert X_t\right]\\ 
& = \left[\widetilde{\log}(P^{h-k})\right]_y\mathbb{E}\left[X_{t+k}\vert X_{t}\right]-\left[\widetilde{\log}(P^{h-k-1})\right]_y\mathbb{E}\left[X_{t+k+1}\vert X_{t}\right]\\
& = \left[\widetilde{\log}(P^{h-k})\right]_yP^{(k)}X_t-\left[\widetilde{\log}(P^{h-k-1})\right]_yP^{k+1}X_t\\
& = \left[\left[\widetilde{\log}(P^{(h-k)})\right]_yP^{k+1}-\left[\widetilde{\log}(P^{h-k-1})\right]_yP^{k}\right]X_t. \\ 
	\end{split}
\end{equation}


\subsubsection{Proof of Corollary 2: Binary Markov Chain as a Special Case of Proposition 5}


The $h$-step transition matrix for the binary Markov Chain is given by: 
\begin{equation*}
	P^{h}= \begin{bmatrix}
		p^{(h)}_{00}&p^{(h)}_{01} \\
		p^{(h)}_{10}&p^{(h)}_{11} \\ 
	\end{bmatrix}= \begin{bmatrix}
		 1-\left[\pi(1-\lambda^{h})\right] &1-\left[\pi+\lambda^{h}(1-\pi)\right]  \\
			 \pi(1-\lambda^{h})& \pi+\lambda^{h}(1-\pi) \\
	\end{bmatrix}
\end{equation*}
For $y=1$, we get: 
\begin{equation*}
	\begin{split}
	 & \left[\left[\widetilde{\log}(P^{h-k})\right]_yP^{k}-\left[\widetilde{\log}(P^{h-k-1})\right]_yP^{k+1}\right]X_t \\ 
		& = \left[\begin{bmatrix}
		 \log p^{(h-k)}_{10} & \log p^{(h-k)}_{11} \\ 
		\end{bmatrix} \begin{bmatrix}
		p^{(k)}_{00}&p^{(k)}_{01} \\
		p^{(k)}_{10}&p^{(k)}_{11} \\ 
	\end{bmatrix}-\begin{bmatrix}
	\log^{(h-k-1)}_{10} & \log p^{(h-k-1)}_{11} \\ 
\end{bmatrix} \begin{bmatrix}
p^{(k+1)}_{00}&p^{(k+1)}_{01} \\
p^{(k+1)}_{10}&p^{(k+1)}_{11} \\ 
\end{bmatrix}\right]\begin{bmatrix}
X_{0,t} \\
X_{1,t} \\
\end{bmatrix}\\ 
& = \begin{bmatrix}
\log p^{(h-k)}_{10}p^{(k)}_{00}+\log p^{(h-k)}_{11}p^{(k)}_{10}-\log p^{(h-k-1)}_{10}p^{(k+1)}_{00}-\log p^{(h-k-1)}_{11}p^{(k+1)}_{10}\\ 
\log p^{(h-k)}_{10}p^{(k)}_{01}+\log p^{(h-k)}_{11}p^{(k)}_{11}-\log p^{(h-k-1)}_{10}p^{(k+1)}_{01}-\log p^{(h-k-1)}_{11}p^{(k+1)}_{11}\\ 
\end{bmatrix}'\begin{bmatrix}
X_{0,t} \\
X_{1,t} \\
\end{bmatrix}\\ 
	\end{split}
\end{equation*}
This expression coincides directly with the formula provided in \eqref{fekd_mc_bin}. For instance, when $Y_t$ is in state 0, $X_t = [1 \ 0]'$ and the above can be simplified to:
\begin{equation*}
	\begin{split}
		& \log p^{(h-k)}_{10}p^{(k)}_{00}+\log p^{(h-k)}_{11}p^{(k)}_{10}-\log p^{(h-k-1)}_{10}p^{(k+1)}_{00}-\log p^{(h-k-1)}_{11}p^{(k+1)}_{10} \\
		=& \log\left[\pi(1-\lambda^k)\right]-\log\left[\pi(1-\lambda^k)\right]\left[\pi(1-\lambda^k)\right]+\log\left[\pi+\lambda^k(1-\pi)\right]\left[\pi(1-\lambda^k)\right] \\
				&- \log\left[\pi(1-\lambda^k)\right]+\log\left[\pi(1-\lambda^k)\right]\left[\pi(1-\lambda^k)\right]-\log\left[\pi+\lambda^k(1-\pi)\right]\left[\pi(1-\lambda^k)\right] \\
				=& \log\left[\frac{1-\lambda^{h-k}}{1-\lambda^{h-k-1}}\right] + \log\left[\frac{\pi+\lambda^{h-k}(1-\pi)}{\pi(1-\lambda^{h-k})}\right]\left[\pi(1-\lambda^k)\right] \\
				& -  \log\left[\frac{\pi+\lambda^{h-k-1}(1-\pi)}{\pi(1-\lambda^{h-k-1})}\right]\left[\pi(1-\lambda^k)\right],
	\end{split}
\end{equation*}
which is equivalent to \eqref{fekd_mc_bin} when $Y_t = 0$.

\subsection{Examples for FELD}

\subsubsection{Proof of Proposition 6: FELD for Dynamic Affine Models}
Since: 
\begin{equation*}
	\mathbb{E}(Y_{t+h}|Y_t) = -\frac{dc}{du}(0) +\left[\frac{da'}{du}(0)\right]^h\left[Y_t + \frac{dc}{du}(0)\right],
\end{equation*}
the left hand side of the FELD is:
\begin{equation*}
	\begin{split}
		& u'\mathbb{E}(Y_{t+h}|Y_t) - a^{\circ h}(u)'Y_t + c(u) -c\left[a^{\circ h}(u)\right] \\
	= & u'\left\{-\frac{dc}{du}(0) +\left[\frac{da'}{du}(0)\right]^h\left[Y_t + \frac{dc}{du}(0)\right]\right\}-a^{\circ h}(u)'Y_t+c(u) -c\left[a^{\circ h}(u)\right]\\ 
	=& \left\{u'\left[\frac{da'}{du}(0)\right]^h - a^{\circ h}(u)'\right\}Y_t - u'\frac{dc}{du}(0)+ u'\left[\frac{da'}{du}(0)\right]^h\left[\frac{dc}{du}(0)\right] + c(u) - c\left[a^{\circ h}(u)\right].\\ 		
	\end{split}
\end{equation*}
Let us now consider the expression: 
\begin{equation*}
	\begin{split}
		& a^{\circ(h-k-1)'}(u)\mathbb{E}(Y_{t+k+1}|Y_t)\\
		=& a^{\circ(h-k-1)'}(u)\left\{-\frac{dc}{du}(0) +\left[\frac{da'}{du}(0)\right]^{k+1}\left[Y_t + \frac{dc}{du}(0)\right]\right\}\\
		= & a^{\circ(h-k-1)'}(u)\left[\frac{da'}{du}(0)\right]^{k+1}Y_t -a^{\circ(h-k-1)'}(u)\left[\frac{dc}{du}(0)\right] +a^{\circ(h-k-1)'}(u)\left[\frac{da'}{du}(0)\right]^{k+1}\left[\frac{dc}{du}(0)\right].
	\end{split}
\end{equation*}
Hence the right hand side of the FELD is: 
\begin{equation*}
	\begin{split}
	&	\sum_{k=0}^{h=1} \left\{a^{\circ(h-k-1)'}(u)\mathbb{E}(Y_{t+k+1}|Y_t)-a^{\circ(h-k)'}(u)\mathbb{E}(Y_{t+k}|Y_t)\right\} + c\left[a^{\circ(h-k-1)}(u)\right]- c\left[a^{\circ(h-k)}(u)\right] \\
		= & \sum_{k=0}^{h-1}\left\{a^{\circ (h-k-1)}(u)'\left[\frac{da'}{du}(0)\right]^{k+1}-a^{\circ (h-k)}(u)'\left[\frac{da'}{du}(0)\right]^{k}\right\}Y_t+\left(a^{\circ (h-k)}(u)'-a^{\circ (h-k-1)}(u)'\right)\left[\frac{dc}{du}(0)\right]\\
	& + \left(a^{\circ (h-k)}(u)'\left[\frac{da'}{du}(0)\right]^k-a^{\circ (h-k-1)}(u)'\left[\frac{da'}{du}(0)\right]^{k+1}\right)\left[\frac{dc}{du}(0)\right] + c\left[a^{\circ (h-k-1)}(u)\right]-c\left[a^{\circ (h-k)}(u)\right], \ \forall u.\\  
	\end{split}
\end{equation*}
as required. \qed

%
%
%
%

\subsubsection{Proof of Corollary 3: FEVD for Dynamic Affine Models}

Let us compute the generic term of the FEVD in (2.4) for an affine process. We have: 
\begin{equation*}
	\mathbb{E}\left(Y_{t+h}\vert I_{t+k+1}\right) = - \frac{dc(0)}{du} + \left(\frac{da'(0)}{du}\right)^{h-k-1}\left(Y_{t+k+1}+\frac{dc(0)}{du}\right).
\end{equation*}
We deduce: 
\begin{equation}
	\begin{split}
		& \mathbb{V}\left[\mathbb{E}\left(Y_{t+h}\vert I_{t+k+1}\right)\vert I_{t+k}\right] \\
		= & \mathbb{V}\left[\left(\frac{\partial a'(0)}{\partial u}\right)^{h-k-1}Y_{t+k+1}\vert I_{t+k}\right]\\
		= & \left(\frac{\partial a'(0)}{\partial u}\right)^{h-k-1}\mathbb{V}\left[Y_{t+k+1}\vert I_{t+k}\right] \left(\frac{\partial a(0)}{\partial u'}\right)^{h-k-1}. \\ 
	\end{split}
\end{equation}
We know that:
\begin{equation*}
	\mathbb{V}\left(Y_{t+1}\vert Y_t\right) = \left[\frac{\partial^2}{\partial u \partial u'}\left(a'(u)Y_t + b(u)\right)\right]_{u=0},
\end{equation*}
where $b(u)=c(u) - c(a(u))$. Therefore we deduce that: 
\begin{equation*}
	\begin{split}
		& \mathbb{V}\left[\mathbb{E}\left(Y_{t+h}\vert I_{t+k+1}\right)\vert I_{t+k}\right]\\
		= & \left[\frac{da'(0)}{du}\right]^{h-k-1}\left[\sum_{j=1}^n \frac{\partial^2a_j(0)}{\partial u\partial u'}Y_{j,t+k+1}+\frac{\partial^2b(0)}{\partial u\partial u'}\right]\left[\frac{da(0)}{du'}\right]^{h-k-1}\\
				= & \sum_{j=1}^n \left\{\left[\frac{da'(0)}{du}\right]^{h-k-1}\frac{\partial^2a_j(0)}{\partial u\partial u'} \left[\frac{da(0)}{du'}\right]^{h-k-1}Y_{j,t+k}\right\}+ \left[\frac{da'(0)}{du}\right]^{h-k-1}\frac{\partial^2b(0)}{\partial u\partial u'} \left[\frac{da(0)}{du'}\right]^{h-k-1}.\\
	\end{split}
\end{equation*}
We deduce the generic term by taking the expectation given $Y_t$. It is equal to: 
\begin{equation*}
\begin{split}
	&	\sum_{j=1}^N \left\{\left(\frac{\partial a'(0)}{du}\right)^{h-k-1}\frac{\partial^2a_j(0)}{\partial u \partial u'}\left(\frac{\partial a'(0)}{du'}\right)^{h-k-1}\left[-\frac{dc(0)}{du_j}+\left(\left(\frac{\partial a'(0)}{\partial u}\right)^k\left(Y_t+\frac{dc(0)}{du}\right)\right)_j\right]\right\}\\ 
	& + \left(\frac{\partial a'(0)}{\partial u}\right)^{h-k-1}\frac{\partial^2b(0)}{\partial u \partial u'}\left(\frac{\partial a(0)}{\partial u'}\right)^{h-k-1}.
\end{split}
\end{equation*}

\subsubsection{Proof of Corollary 5: Markov Chains}

The generic term on the right hand side of the FELD is:
\begin{equation*}
	\begin{split}
	&	\mathbb{E}\left[\widetilde{\log}\left(\exp(u)P^{h-k}\right)X_{t+k}-\widetilde{\log}\left(\exp(u)P^{h-k-1}\right)X_{t+k+1}\big\vert X_t\right] \\
		= & \widetilde{\log}\left(\exp(u)P^{h-k}\right)\mathbb{E}\left[X_{t+k}\vert X_t\right]-\widetilde{\log}\left(\exp(u)P^{h-k-1}\right)\mathbb{E}\left[X_{t+k+1}\vert X_t\right]\\  
		= & \widetilde{\log}\left(\exp(u)P^{h-k}\right)P^{k}X_t-\widetilde{\log}\left(\exp(u)P^{h-k-1}\right)P^{k+1}X_t\\ 
		= & \left[\widetilde{\log}\left(\exp(u)P^{h-k}\right)P^{k}-\widetilde{\log}\left(\exp(u)P^{h-k-1}\right)P^{k+1}\right]X_t.\\ 
	\end{split}
\end{equation*}

\subsubsection{Proof of Corollary 6: INAR(1)}

The conditional log-Laplace transform of an INAR(1) process at horizon 1 is:
\begin{equation*}
	\log \left[\Psi(u,1\vert I_t)\right]=	\lambda\left[1-\exp (-u)\right]+Y_{t}\log\left[1-p+p\exp (-u)\right], 
\end{equation*}
which is affine in $Y_t$ and can be written in the form of equation \eqref{cllt}. In particular, we have [Darolles et al. (2004)]:
\begin{equation*}
	\begin{split}
		a(u) & = -\log\left[1-p+p\exp (-u)\right], \\
		c(u) & = \frac{-\lambda}{1-p}\left[1-\exp (-u)\right]. \\
	\end{split}
\end{equation*}
We deduce that: 
\begin{equation*}
	\begin{split}
		\frac{da}{du}(0) &= \frac{p}{p-(p-1)\exp(0)} = p,  \\
		\frac{dc}{du}(0) &= -\frac{\lambda\exp(0)}{1-p} = -\frac{\lambda}{1-p}.\\ 
 	\end{split}
\end{equation*}
Hence, we have: 
\begin{equation*}
	\mathbb{E}\left[Y_{t+h}\vert Y_t\right]= p^hY_t + \frac{\lambda }{1-p}(1-p^h).
\end{equation*}

It is known that the moment generating function of the INAR(1) model at horizon $h$ has the same form as the moment generating function with parameters $(\lambda,p)$ replaced by $(\lambda \frac{1-p^h}{1-p},p^h)$. Therefore, we get: 
\begin{equation*}
	\begin{split}
		a^{\circ (h)}(u) = a[a^{\circ (h-1)}(u)] = -\log\left[1-p^h+p^h\exp (-u)\right], \\
	\end{split}
\end{equation*}
and: 
\begin{equation*}
	c\left[a^{\circ (h)}(u) \right]= \frac{\lambda}{1-p} p^h \left[1-\exp (-u)\right].
\end{equation*}
Combining these results and applying Proposition 6, the left hand side of the FELD is:
\begin{equation*}
\begin{split}
	& \left\{p^hu+\log\left[1-p^h+p^h\exp (-u)\right]\right\}Y_t	+\frac{\lambda}{1-p}u-\frac{\lambda}{1-p}p^hu+\frac{\lambda}{1-p}\left[1-\exp (-u)\right]- \frac{\lambda}{1-p} p^h \left[1-\exp(-u)\right]\\
=	& \left\{p^hu+\log\left[1-p^h+p^h\exp (-u)\right]\right\}Y_t+\frac{\lambda}{1-p}\left[\left(1-p^h\right)\left(u-1+\exp(-u)\right)\right]. \\
\end{split}
\end{equation*}
By using the expansions of $a^{\circ h}(u)$, $c\left[a^{\circ h}(u)\right]$ and of the prediction $\mathbb{E}(Y_{t+h}\vert Y_t)$, the right hand side is: 
\begin{equation*}
	\begin{split}
 & \sum_{k=0}^{h-1}\left\{p^k\log\left[1-p^{h-k}+p^{h-k}\exp (-u)\right]-p^{k+1}\log\left[1-p^{h-k-1}+p^{h-k-1}\exp (-u)\right]\right\}Y_t\\
& +\lambda \frac{(1-p^k)}{1-p}\log \left(1-p^{h-k}+p^{h-k}\exp(u)\right)-\lambda \frac{(1-p^{k+1})}{1-p}\log \left(1-p^{h-k-1}+p^{h-k-1}\exp(u)\right)\\ 
& +\lambda \left[1-\exp(u)\right]p^{h-k-1}.\\ 
	\end{split}
\end{equation*}

\subsubsection{Proof of Corollary 7: ARG(1)}

For the ARG(1) process we know that:
\begin{equation*}
	\begin{split}
		a(u) = & \frac{\beta u}{1+u}, \\
		c(u) = & -\delta \log\left[1+\frac{u}{1-\beta}\right],\\
		a^{\circ h}(u) =& \frac{\beta^h u}{\left[1+\frac{1-\beta^h}{1-\beta}u\right]}. \\ 
 	\end{split}
\end{equation*}
It is easily checked that:
\begin{equation*} 
	\frac{da}{du}(0) = \left[\frac{\beta}{(1+u)^2}\right]_{u=0} = \beta.
\end{equation*}
 
Thus applying Proposition 5:
\begin{equation*}
	\begin{split}
		\alpha(h,u)&= u\left[\frac{da}{du}(0)\right]^h-a^{\circ h}(u)\\
		& = \beta^hu -\frac{\beta^h u}{\left[1+\frac{1-\beta^h}{1-\beta}u\right]}\\
		& = \beta^hu\left[1-\frac{1}{\left(1+\frac{1-\beta^h}{1-\beta}u\right)}\right]. \\  
	\end{split}
\end{equation*}
Similarily: 
\begin{equation*}
	\begin{split}
		a(h,k,u) & = a^{\circ (h-k-1)}(u) \left[\frac{da}{du}(0)\right]^{k+1}-a^{\circ (h-k)}(u) \left[\frac{da}{du}(0)\right]^{k}\\
		& = \frac{\beta^{h-k-1} u}{\left[1+\frac{1-\beta^{h-k-1}}{1-\beta}u\right]}\beta^{k+1}-\frac{\beta^{h-k} u}{\left[1+\frac{1-\beta^{h-k}}{1-\beta}u\right]}\beta^{k}\\
		& = \beta^hu\left[\frac{1}{\left(1+\frac{1-\beta^{h-k-1}}{1-\beta}u\right)}-\frac{1}{\left(1+\frac{1-\beta^{h-k}}{1-\beta}u\right)}\right].\\ 
	\end{split}
\end{equation*}
To get the terms $\beta(h,u)$ and $\beta(h,k,u)$, note:
\begin{equation*}
	\frac{dc}{du}(0) = -\delta\left[\frac{1}{u-\beta+1}\right]_{u=0}=-\frac{\delta}{1-\beta},
\end{equation*}
and,
\begin{equation*}
	\begin{split}
		c\left[a^{\circ h}(u)\right] & = -\delta \log \left[1+\frac{a^{\circ h}(u)}{1-\beta}\right]\\
		& = -\delta \log \left[1+\frac{\beta^hu}{(1-\beta)\left(1+\frac{1-\beta^{h-k}}{1-\beta}u\right)}\right]\\
		& = -\delta \log \left[1+\frac{\beta^hu}{(1-\beta)\left(1-\beta+(1-\beta^h)u\right)}\right].\\
	\end{split}
\end{equation*}
Hence, we apply Proposition 5 to obtain: 
\begin{equation*}
\begin{split}
		\beta(h,u)&= - u\frac{dc}{du}(0) + u\left[\frac{da}{du}(0)\right]^h\frac{dc}{du}(0)+c(u)-c\left[a^{\circ h}(u)\right]\\
		& = \frac{\delta u}{1-\beta}-\frac{\beta^hu\delta}{1-\beta}-\delta\log\left[1+\frac{u}{1-\beta}\right]+\delta\log\left[1+\frac{\beta^hu}{1-\beta+(1-\beta^h)u}\right].
\end{split}
\end{equation*}
Similarly,
\begin{equation*}
	\begin{split}
		\beta(h,k,u)= &  \left(a^{\circ(h-k)}(u)-a^{\circ(h-k-1)(u)}\right)\left(\frac{dc}{du}(0)\right) \\
		&+\left[a^{\circ(h-k)}(u)\left(\frac{da}{du}(0)\right)^k-a^{\circ(h-k-1)(u)}\left(\frac{da}{du}(0)\right)^{k+1}\right]\left(\frac{dc}{du}(0)\right)\\
		& + c\left(a^{\circ (h-k-1)}(u)\right)-c\left(a^{\circ (h-k)}(u)\right) \\
		=& \left[\frac{\beta^{h-k} u}{\left(1+\frac{1-\beta^{h-k}}{1-\beta}u\right)}-\frac{\beta^{h-k-1} u}{\left(1+\frac{1-\beta^{h-k-1}}{1-\beta}u\right)}\right]\left(\frac{-\delta}{1-\beta}\right) \\ 
		& + \left[\frac{\beta^{h} u}{\left(1+\frac{1-\beta^{h-k}}{1-\beta}u\right)}-\frac{\beta^{h} u}{\left(1+\frac{1-\beta^{h-k-1}}{1-\beta}u\right)}\right]\left(\frac{-\delta}{1-\beta}\right) \\ 
		& + \delta \log\left[1+\frac{\beta^{h-k-1}u}{1-\beta+(1-\beta^{h-k-1}u)}\right]-\delta \log\left[1+\frac{\beta^{h-k}u}{1-\beta+(1-\beta^{h-k}u)}\right].
	\end{split}
\end{equation*}
 
\subsubsection{Proof for the Wishart Process}

Since the Wishart process concerns stochastic symmetric positive definite matrices, the conditional Laplace transform is usually written with matrix arguments [see Gourieroux, Jasiak and Sufana (2009)]. It is given by:
\begin{equation*}
\begin{split}
		\Psi(\Gamma,h\vert I_t ) & = \mathbb{E}\left[\exp(-\Tr(\Gamma Y_{t+h}))\vert Y_t\right] \\ 
		& = \frac{\exp\left[-\Tr\left(\left(M^{h'}\right)\Gamma(Id+2\Sigma_h \Gamma)^{-1}M^hY_t\right)\right]}{\det(Id+2\Sigma_h\Gamma)^{k/2}}, \\
\end{split}
\end{equation*}
where $\Sigma_h=\Sigma + M\Sigma M' + ... + M^{h-1}\Sigma (M^{h-1})'$. Therefore we get: 
\begin{equation*}
	\begin{split}
		&\log \Psi (\Gamma,h-k \vert I_{t+k}) - \log \Psi\left(\Gamma,h-k-1\vert I_{t+k+1}\right) \\
		=& \Tr \left\{(M^{h-k-1})'\Gamma(Id+2\Sigma_{h-k-1}\Gamma)^{-1}M^{h-k-1}Y_{t+k+1}\right\} \\
		& -\Tr \left\{(M^{h-k})'\Gamma(Id+2\Sigma_{h-k}\Gamma)^{-1}M^{h-k}Y_{t+k}\right\}\\
		& - \log\left[\det\left(Id+2\Sigma_{h-k}\Gamma\right)^{K/2}\right] \\ 
		& + \log\left[\det\left(Id+2\Sigma_{h-k-1}\Gamma\right)^{K/2}\right].\\ 
	\end{split}
\end{equation*}
Conditioning on $Y_t$ in this equation, noting that  $\mathbb{E}(Y_{t+h}\vert Y_t) = M^hY_t(M^h)'+K\Sigma_h$, the generic term in the FELD decomposition \eqref{FELD} is: 
\begin{equation*}
	\begin{split}
		& \Tr \left\{(M^{h})'\Gamma\left[(Id+2\Sigma_{h-k-1}\Gamma)^{-1}-(Id+2\Sigma_{h-k}\Gamma)^{-1}\right]M^h Y_t\right\} \\ 
		+ & \left\{\Tr\left[(M^{h-k})'\Gamma(Id+2\Sigma_{h-k}\Gamma)^{-1}M^{h-k}K\Sigma_{k}\right]\right\} \\ 
		- & \left\{\Tr\left[(M^{h-k-1})'\Gamma(Id+2\Sigma_{h-k-1}\Gamma)^{-1}M^{h-k-1}K\Sigma_{k+1}\right]\right\} \\ 
		+ & \frac{K}{2}\log \left[\frac{\det\left(Id+2\Sigma_{h-k}\Gamma\right)}{\det\left(Id+2\Sigma_{h-k-1}\Gamma\right)}\right].
	\end{split}
\end{equation*}

\subsubsection{Conditional Laplace Transform for Bivariate NBAR at Horizon 1}

\begin{equation*}
	\begin{split}
		& \mathbb{E}\left[\exp(-u'Y_{t+1})|Y_{t}\right]\\
		= & \mathbb{E}\left[\mathbb{E}(\exp(-u'Y_{t+1})|\underline{Y}_t,\underline{X}_{t+1},\underline{Z}_{t+1})|Y_{t}\right]\\
		= & \mathbb{E}\left[\exp(-(\alpha_1Z_{t+1}+\beta_1X_{1,t+1})(1-\exp(-u_1)))\exp(-(\alpha_2Z_{t+1}+\beta_2X_{2,t+1})(1-\exp(-u_2)))|Y_t\right]\\
				= & \mathbb{E}\left\{\exp\left[-\alpha_1Z_{t+1}(1-\exp(-u_1))-\alpha_2Z_{t+1}(1-\exp(-u_2))\right]\right.\\
				& \left.\times\exp\left[\beta_1X_{1,t+1}(1-\exp(-u_1))\right]\times\exp\left[\beta_2X_{2,t+1}(1-\exp(-u_2))\right]|Y_t\right\}\\
				= & \frac{1}{\left[1+\alpha_1(1-\exp(-u_1))+\alpha_2(1-\exp(-u_2))\right]^{\delta+\sigma_1Y_{1,t}+\sigma_2Y_{2,t}}} \\
				& \times \frac{1}{\left[1+\beta_1(1-\exp(-u_1))\right]^{\delta_1+Y_{1,t}}}\times \frac{1}{\left[1+\beta_2(1-\exp(-u_2))\right]^{\delta_2+Y_{2,t}}}\\  
				= & \frac{1}{\left[1+\beta_1(1-\exp(-u_1))\right]^{Y_{1,t}}} \times \frac{1}{\left[1+\alpha_1(1-\exp(-u_1))+\alpha_2(1-\exp(-u_2))\right]^{\sigma_1Y_{1,t}}} \\
				 & \times \frac{1}{\left[1+\beta_2(1-\exp(-u_2))\right]^{Y_{2,t}}}\times \frac{1}{\left[1+\alpha_1(1-\exp(-u_1))+\alpha_2(1-\exp(-u_2))\right]^{\sigma_2Y_{2,t}}} \\
				 				 & \times \frac{1}{\left[1+\alpha_1(1-\exp(-u_1))+\alpha_2(1-\exp(-u_2))\right]^{\delta}} \\
				 				 & \times \frac{1}{\left[1+\beta_1(1-\exp(-u_1))\right]^{\delta_1}}\times \frac{1}{\left[1+\beta_2(1-\exp(-u_2))\right]^{\delta_2}}\\
				= &\exp\left[-a_1(u_1,u_2)Y_{1,t}-a_2(u_1,u_2)Y_{2,t}-b(u_1,u_2)\right],
	\end{split}
\end{equation*}
where: 
\begin{equation*}
	\begin{split}
		a_1(u_1,u_2)= &  \log\left[1+\beta_1(1-\exp(-u_1))\right]+\sigma_1\log[1+\alpha_1(1-\exp(-u_1))+\alpha_2(1-\exp(-u_2))],\\
		a_2(u_1,u_2)= &  \log\left[1+\beta_2(1-\exp(-u_2))\right]+\sigma_2\log[1+\alpha_1(1-\exp(-u_1))+\alpha_2(1-\exp(-u_2))],\\
		b(u_1,u_2) =&  \delta_1\log[1+\beta_1(1-\exp(-u_1))]+\delta_2\log[1+\beta_2(1-\exp(-u_2))]\\
		& + \delta_3\log[1+\alpha_1(1-\exp(-u_1))+\alpha_2(1-\exp(-u_2))].\\ 
	\end{split}
\end{equation*}

\subsubsection{VAR(1) Representation for the Bivariate NBAR Linear Prediction Formula}

\begin{equation*}
	\begin{split}
		&\mathbb{E}\left[Y_{t}|Y_{t-1}\right]\\
		= & \mathbb{E}\left[\mathbb{E}(Y_t|\underline{Y}_{t-1},\underline{X}_{t},\underline{Z}_{t})|Y_{t-1}\right]\\ 
		= & \mathbb{E}\left[\begin{bmatrix}
			\alpha_1\\
			\alpha_2
		\end{bmatrix}Z_{t} + \begin{bmatrix}
		\beta_1 X_{1,t+1}\\
		\beta_2 X_{2,t+1}
		\end{bmatrix}\bigg\vert Y_{t-1}\right]\\
		= & \begin{bmatrix}
			\alpha_1 \\
			\alpha_2 \\
		\end{bmatrix} \left[\delta_+\sigma_1Y_{1,t-1}+\sigma_2Y_{2,t-1}\right]+ \begin{bmatrix}
			\beta_1 \left(\delta_1+Y_{1,t-1}\right)\\
			\beta_2 \left(\delta_2+Y_{2,t-1}\right)
		\end{bmatrix} \\
	= & \begin{bmatrix}
		\alpha_1\delta + \beta_1\delta\\
		\alpha_2\delta + \beta_2\delta\\ 
	\end{bmatrix}+\begin{bmatrix}
	\alpha_1\sigma_1+\beta_1  &\alpha_1\sigma_2 \\
	\alpha_2\sigma_1 &  \alpha_2\sigma_2 +\beta_2 \\
	\end{bmatrix}\begin{bmatrix}
	Y_{1,t-1}\\
	Y_{2,t-1}
	\end{bmatrix}.
		\end{split}
\end{equation*}

%

\newpage

\section{Online Appendices}

\subsection{Asymptotic Confidence Bands on the FELD}

Suppose $\hat{\theta}_T$ is a consistent estimator of $\theta$ and is asymptotically normal, that is: 
\begin{equation*}
	\sqrt{T}\left(\hat{\theta}_T - \theta\right) \xrightarrow{d} N(0,V),
\end{equation*}
where $\xrightarrow{d}$ denotes the convergence in distribution. Then by the Delta method, we have, for given $h$: 
\begin{equation*}
\sqrt{T}\left[\text{vec}\left(\gamma(k,h\vert u,y;\hat{\theta}_t)\right)-\text{vec}\left(\gamma(k,h\vert u,y;\theta_0)\right)\right] \xrightarrow{d} N\left(0,\frac{\partial \Gamma(u,y,\theta_0)}{\partial\theta'}V\frac{\partial \Gamma(u,y,\theta_0)'}{\partial\theta}\right), 
\end{equation*}
where $\frac{\partial \Gamma(u,y;\theta_0)}{\partial\theta'}=\left[\frac{\partial\gamma(1,h\vert u,y;\theta_0)}{\partial \theta},...,\frac{\partial\gamma(h-1,h\vert u,y;\theta_0)}{\partial \theta}\right]$. Hence, the asymptotic distribution of $\text{vec} \left[\gamma(k,h\vert y,u;\hat{\theta}_t)\right]$ can be approximated using a Gaussian with mean $\gamma(k,h\vert y,u;\theta_0)$ and variance $\Omega_T(u,y) = \frac{1}{T}\left[\frac{\partial \Gamma(u,y,\hat{\theta}_T)}{\partial\theta'}\hat{V}\frac{\partial \Gamma(u,y,\hat{\theta}_T)}{\partial\theta}'\right]$, where $\hat{V}_T$ is a consistent estimator of $V$. Then the asymptotic distribution of $\hat{\gamma}(h|u,y)=\gamma(h|u,y;\hat{\theta}_T)$ can be approximated by the Gaussian distribution with mean $\gamma(h|u,y;\theta_0)$ and variance $e'\Omega_T(u,y)e$, where $e$ is the vector with unitary components. \\

%

\textbf{Remark:} For dynamic affine models, the closed form expressions of the components in the FELD contain terms of the form $a^{\circ k}(u;\theta)$, when $\theta$ is explicitly written. Since $a^{\circ k}(u;\theta)=a\left[a^{\circ (h-1)}(u;\theta);\theta\right]$, we get:
\begin{equation*}
	\frac{\partial a^{\circ k}(u;\theta)}{\partial \theta'} = \frac{\partial a}{\partial \theta'}\left[ a^{\circ (k-1)}(u;\theta);\theta\right] + \frac{\partial a}{\partial u'} \left[a^{\circ (k-1)}(u;\theta);\theta\right]\frac{\partial \left[a^{\circ (k-1)}(u;\theta);\theta\right]}{\partial \theta}, \ \forall k.
\end{equation*}
The computation this of the derivative $\frac{\partial \gamma}{\partial \theta}$ can be simplified by applying recursively this rule.  

\subsection{Negative Binomial Autoregressive Process}

\subsubsection{Proof of Corollary 9: FELD for Univariate NBAR Processes}

Since the NBAR is a dynamic affine process, we may apply Propostion 6 directly to obtain the FELD. From the conditional Laplace transform in \eqref{lptunbar} we get: 
\begin{equation*}
	a(u) = \log \left[1+\beta  c(1-\exp(-u))\right],
\end{equation*}
and 
\begin{equation*}
	\begin{split}
		a^{\circ h }(u) & =-\log\left[1+\beta c_{h-1}(1-\exp(-u))\right]+\log\left[1+\beta c_{h}(1-\exp(-u))\right],\\
		b^{\circ h }(u) & = c(u) - c\left[a^{\circ h }(u)\right] =- \delta \log\left[1+\beta c_{h}(1-\exp(-u))\right].\\
	\end{split}
\end{equation*}
The unconditional Laplace transform is given by \eqref{unclptunnbar} is:
\begin{equation*}
	c(u) = -\delta\log\left[1+\beta c(1-\exp(-u))\right].
\end{equation*}
Hence, we deduce that:
\begin{equation*}
	\frac{da}{du}(0) = \frac{\beta c}{\beta c \left(\exp(u)-1\right)+\exp(u)}\bigg\vert_{x=0} = \beta c = \rho ,
\end{equation*}
and:
\begin{equation*}
	\frac{dc}{du}(0) = \frac{-\delta \beta c}{\beta c \left(\exp(x)-1\right)+\exp(x)}\bigg\vert_{x=0} = -\delta \beta c = -\delta \rho.
\end{equation*}
Applying Propostion 6, the LHS of the FELD is given by:
\begin{equation*}
	\begin{split}
	&	\left\{u'\left[\frac{da'}{du}(0)\right]^h - a^{\circ h}(u)'\right\}Y_t - u'\frac{dc}{du}(0)+ u'\left[\frac{da'}{du}(0)\right]^h\left[\frac{dc}{du}(0)\right] + c(u) - c\left[a^{\circ h}(u)\right] \\ 
	= & \left\{u\rho^h+\log\left[1+\beta c_{h-1}(1-\exp(-u))\right]-\log\left[1+\beta c_{h}(1-\exp(-u))\right]\right\}Y_t\\ 
	& +\delta \rho u - \delta \rho^{h+1} u - \delta \log\left[1+\beta c_{h}(1-\exp(-u))\right]. \\
	\end{split}
\end{equation*}
Likewise, the RHS of the FELD is given by:
\begin{equation*}
\begin{split}
& \sum_{k=0}^{h-1}\left\{a^{\circ (h-k-1)}(u)'\left[\frac{da'}{du}(0)\right]^{k+1}-a^{\circ (h-k)}(u)'\left[\frac{da'}{du}(0)\right]^{k}\right\}Y_t+\left(a^{\circ (h-k)}(u)'-a^{\circ (h-k-1)}(u)'\right)\left[\frac{dc}{du}(0)\right]\\
& + \left(a^{\circ (h-k)}(u)'\left[\frac{da'}{du}(0)\right]^k-a^{\circ (h-k-1)}(u)'\left[\frac{da'}{du}(0)\right]^{k+1}\right)\left[\frac{dc}{du}(0)\right] + c\left[a^{\circ (h-k-1)}(u)\right]-c\left[a^{\circ (h-k)}(u)\right]\\
= & \sum_{k=0}^{h-1}\left\{\rho^{k+1}\log\left[\frac{1+\beta c_{h-k-1}(1-\exp(-u))}{1+\beta c_{h-k-2}(1-\exp(-u))}\right]-\rho^k\log\left[\frac{1+\beta c_{h-k}(1-\exp(-u))}{1+\beta c_{h-k-1}(1-\exp(-u))}\right]\right\}Y_t\\
& -\delta \rho\left\{\log\left[\frac{1+\beta c_{h-k}(1-\exp(-u))}{1+\beta c_{h-k-1}(1-\exp(-u))}\right]-\log\left[\frac{1+\beta c_{h-k-1}(1-\exp(-u))}{1+\beta c_{h-k-2}(1-\exp(-u))}\right]\right\} \\
& -\delta \rho \left\{\rho^{k}\log\left[\frac{1+\beta c_{h-k}(1-\exp(-u))}{1+\beta c_{h-k-1}(1-\exp(-u))}\right]-\rho^{k+1}\log\left[\frac{1+\beta c_{h-k-1}(1-\exp(-u))}{1+\beta c_{h-k-2}(1-\exp(-u))}\right]\right\}\\
& - \delta \log\left[\frac{1+\beta c_{h-k}(1-\exp(-u))}{1+\beta c_{h-k-1}(1-\exp(-u))}\right].\\
\end{split}
\end{equation*}

\subsubsection{Extension to $K$-dimensions}

The $J$ dimensional NBAR process $(Y_t)$ satisfies a nonlinear state space model with the latent state variables $X_t = (X_{1,t},...,X_{J,t})$, $Z_t=(Z_{1,t},...,Z_{J,t})$ characterized by the measurement and state equations below:
\begin{itemize}
	\item \textbf{Measurement Equation -} They define the conditional distribution of $Y_{t+1}$ given $\underline{Z}_{t+1}$,$\underline{X}_{t+1}$,$\underline{Y}_{t}$. It is assumed that $Y_{1,t+1},...,Y_{J,t+1}$ are conditionally independent with Poisson distribution $\mathcal{P}(X_{j,t+1}+Z_{j,t+1})$ for $j=1,...,J$.
	\item \textbf{Transition Equations -} They define the conditional distribution of $X_{t+1}$ and $Z_{t+1}$ given $\underline{Z}_{t}$,$\underline{X}_{t}$,$\underline{Y}_{t}$. The conditional intensity of the Poisson distribution has idiosyncratic components $Z_{j,t+1}$ for $j=1,...,J$ and cross-dependent components $X_{j,t+1}$. It is assumed that these variables are conditionally independent where:
	\begin{enumerate}
		\item $Z_{j,t+1}$ follows a gamma distribution $\gamma(\delta_j+\beta'_jY_t,c_j)$, $j=1,...,J$ with $\delta_j+\beta_j'Y_t$, $\delta_j>0$, $\beta_j>0$ the path dependent degree of freedom and $c_j>0$ the scale parameter. 
		\item The cross-dependent intensities $X_{t+1}$ follow a Wishart Gamma distribution with a conditional Laplace transform: 
				\begin{equation*}
						\mathbb{E}\left[\exp(-s_1X_{1,t+1}-...-s_JX_{J,t+1})\bigg\vert Y_t\right] = \frac{1}{\left\{\det\left[I_d+2\textup{diag}(s)\Sigma\right]\right\}^{\delta_0/2+\alpha'Y_t}},
					\end{equation*}
			where the components of $s=(s_1,...,s_J)$ are positive. The parameterization involves a positive definite matrix $\Sigma$ to create the cross-sectional dependence and a path dependent degree of freedom $\delta_0/2 + \alpha'Y_{t}$, where $\delta_0$ and $\alpha_j>0$, $j=1,...,J$. 
	\end{enumerate}
\end{itemize}
This NBAR model can be applied to the four series DISC, HACK, INSD and ELET with $J=4$, or pairwise by couple of series $(J=2)$ or to each single series $(J=1)$. The causal representation in the case of $J=2$ is given by:\\

\begin{equation*}
	\begin{tikzcd}
		&Z_{1,t} \arrow[rd,shift left = 0.5ex]& &Z_{1,t+1}\arrow[rd,shift left = 1.5ex]\\ 
		\left.\begin{matrix}
			Y_{1,t-1}\\
			Y_{2,t-1}
		\end{matrix}\right]  \arrow[r] \arrow[ru] \arrow[rd] & \left[\begin{matrix}
		X_{1,t}\\
		X_{2,t}
	\end{matrix}\right.\arrow[r,shift right=1.5ex]\arrow[r,shift left=1.5ex] & \left.\begin{matrix}
	Y_{1,t}\\
	Y_{2,t}
\end{matrix}\right] \arrow[r] \arrow[ru] \arrow[rd] & \left[\begin{matrix}
	X_{1,t+1} \\
	X_{2,t+1} \end{matrix}\right. \arrow[r,shift right=1.5ex]\arrow[r,shift left=1.5ex] & \left.\begin{matrix}
	Y_{1,t+1}\\
	Y_{2,t+1}
\end{matrix}\right.\\
		& Z_{2,t} \arrow[ru,shift right = 0.5ex] & & Z_{2,t+1}\arrow[ru,shift right = 1.5ex]\\ 
	\end{tikzcd} 
\end{equation*}
In Lu et al. (2024), the nonlinear prediction formula is given by the probability generating function $\mathbb{E}(u_1^{Y_{1,t+1}}...u_1^{Y_{K,t+1}}|Y_t)$. To facilitate the construction of the FELD, we instead express the nonlinear prediction formula by means of the log-Laplace transform:
\begin{lemma}
	The log-Laplace transform of the multivariate NBAR process is given by:
	\begin{equation*}
		\begin{split}
			&\mathbb{E}\left[\exp(-u_1Y_{1,t+1}-...-u_JY_{J,t+1})\vert Y_t\right]  \\
			=& \mathbb{E}\left\{	\mathbb{E}\left[\exp(-u_1Y_{1,t+1}-...-u_JY_{J,t+1}) \vert X_{t+1}, Z_{t+1}\right]\bigg \vert Y_t\right\} \\
			=& \mathbb{E}\left\{\exp\left[-(1-\exp(-u_1))(X_{1,t+1}+Z_{1,t+1})-...-(1-\exp(-u_J))(X_{J,t+1}+Z_{J,t+1})\right]\bigg \vert Y_t\right\}\\
			=& \mathbb{E}\left\{\exp\left[\sum_{j=1}^{J}-(1-\exp(-u_j))Z_{j,t+1}\right]\exp\left[\sum_{j=1}^{J}-(1-\exp(-u_j))X_{j,t+1}\right]\bigg\vert Y_t\right\}\\ 
			= & \left[\prod_{j=1}^{J}\frac{1}{\left[1+c_j(1-\exp(-u_j))\right]^{\delta_j+\beta_j'Y_{t}}}\right]\times \frac{1}{\left\{\det\left[Id+2\textup{diag}((1-\exp(-u_1)),...,(1-\exp(-u_J)))\Sigma\right]\right\}^{\delta_0/2+\alpha'Y_t}}\\ 
			= & \exp\left[-a(u_1,...,u_J)'Y_t-b(u_1,...,u_J)\right],\\
		\end{split}
	\end{equation*}
	where the expressions for $a(u_1,...,u_J)=(a_1,...,a_J)$ and $b(u_1,...,u_J)$ are given by: 
	\begin{equation*}
		\begin{split}
			a_i(u_1,...,u_J) =& \sum_{j=1}^{J}\beta_{i,j} \log\left[1+c_i(1-\exp(-u_i))\right]\\\
			& +\alpha_i\log\left\{\det\left[Id+2\textup{diag}((1-\exp(-u_1)),...,(1-\exp(-u_J)))\Sigma\right]\right\}, \ \ \textup{for} \ j=1,...,J,\\
			b(u_1,...,u_J) & = \sum_{j=1}^J\delta_j\log\left[1+c_j(1-\exp(-u_j))\right]\\
			& +\frac{\delta_0}{2}\log\left\{\det\left[Id+2\textup{diag}((1-\exp(-u_1)),...,(1-\exp(-u_J)))\Sigma\right]\right\},\\
		\end{split}
	\end{equation*}
\end{lemma}
Since this is dynamic affine, the $h$-step ahead log-Laplace transform can be recursively defined as follows:
\begin{equation*}
	\begin{split}
		a^{(h)}(u_1,...,u_J)& = a_i(a_1^{(h-1)},...,a_J^{(h-1)}), \\
		b^{(h)}(u_1,...,u_J)& = b(a_1^{(h-1)},...,a_J^{(h-1)}). \\
	\end{split}
\end{equation*}
The FELD can then be generated using this recursive formula, in a similar manner to the bivariate case, described in Section 7.3.3.

\end{document}